\documentclass{article}
\usepackage[onecolumn]{emulateapj}

\def\kms{km s$^{-1}$}

\def\mh{M_{\bullet}}

\def\mf{m_f}
\def\nf{n_f}
\def\ff{f_f}
\def\sf{\sigma_f}
\def\rf{\rho_f}
\def\vf{v_f}
\def\xf{x_f}
\def\pf{p_f}
\def\m12{M_{12}}
\def\msun{M_{\odot}}

\def\vp{v_{\parallel}}
\def\vt{v_{\perp}}
\def\vb{V_{\rm bin}}
\def\df{\langle\Delta\vp\rangle}
\def\dvp{\langle\Delta\vp^2\rangle}
\def\dvt{\langle\Delta\vt^2\rangle}
\def\ac{A_{\cal C}}
\def\bc{B_{\cal C}}
\def\cc{C_{\cal C}}
\def\dc{D_{\cal C}}
\def\ec{E_{\cal C}}
\def\fc{F_{\cal C}}
\def\beq{\begin{equation}}
\def\eeq{\end{equation}}
\def\fun#1#2{\lower3.6pt\vbox{\baselineskip0pt\lineskip.9pt
  \ialign{$\mathsurround=0pt#1\hfil##\hfil$\crcr#2\crcr\sim\crcr}}}
\def\lap{\mathrel{\mathpalette\fun <}}
\def\gap{\mathrel{\mathpalette\fun >}}

\lefthead{Merritt}
\righthead{Brownian Motion}

\begin{document}

\title{Brownian Motion of a Massive Binary}
\author{David Merritt}
\affil{Department of Physics and Astronomy}

\begin{abstract}

The dynamical friction and diffusion coefficients
are derived for a massive binary that moves against a uniform
background of stars.
The random impulses exerted on the binary's center of mass by the
field stars are greater than those exerted on a point particle due 
to inelastic scattering.
The frictional force acting on the binary is less than that
acting on a point particle due to randomization of the trajectories of
field stars that pass near the binary.
Both effects tend to increase the random motion of
a binary compared with that of a point mass.
If the maximum effective impact parameter for gravitational 
encounters is comparable to the radius of gravitational influence 
of the binary, 
its Brownian velocity can be increased by a modest factor ($\lap 2$)
compared with that of a single particle.
This condition is probably fulfilled in the case of binary 
supermassive black holes in galactic nuclei.

\end{abstract}

\section{Introduction}

Brownian motion, the irregular motion exhibited by a heavy particle
immersed in a fluid of lighter particles, 
is usually modelled by the Langevin equation,
\beq
{d{\bf v}\over dt} = -A{\bf v} + {\bf F}(t).
\eeq
Here ${\bf v}$ denotes the velocity of the heavy particle, 
$-A{\bf v}$ is the frictional force from the fluid,
and ${\bf F}(t)$ is the fluctuating force that arises from collisions
between the heavy particle and the fluid particles.
The precise form of ${\bf F}(t)$ is typically not known but its 
statistical properties are well defined; 
for instance, the mean value of ${\bf F}$ is zero and its fluctuations
generally occur on a time scale that is short compared to the characteristic
time over which ${\bf v}$ varies.
When these conditions are met, 
the evolution of the single-particle distribution function $f$ is
described by the Fokker-Planck equation (e.g. \cite{van92});
the dynamical friction coefficient that appears in that equation
is just $A$,
and the diffusion coefficient is related to the amplitude of 
${\bf F}$ via the requirement that the steady-state motion be in energy 
equipartition with the fluid at temperature $T$ (\cite{ein55}).

This paper is concerned with the Brownian motion of a massive {\it binary},
consisting of two point particles whose center of mass moves 
irregularly due to random gravitational encounters with field stars.
The Brownian motion of a binary differs from that of a single particle
because the kinetic energy of a field star is not conserved following
a close interaction with the binary.
If the binary's mass is much greater than that of a field star,
most binary-field star interactions will extract energy from
the binary (\cite{heg75}),
increasing the field star's kinetic energy and hence
the recoil velocity of the binary's center of mass.
This ``super-elastic scattering'' will give the binary 
a larger random velocity than expected for a
point particle in energy equipartition with background stars, 
or
\beq
\langle v^2\rangle = {\mf\over M}\langle\vf^2\rangle,
\label{equipart}
\eeq
where $M$ and $\mf$ are the mass of the heavy particle and of a field 
star respectively,
and $\langle\vf^2\rangle$ is the mean square velocity of the field
stars.

A motivation for the work presented here is the likely existence
of binary black holes.
There are at least two environments where binary black holes are
expected to form.
In globular clusters, stars with initial masses exceeding $\sim 20\msun$
should comprise $\sim 0.1\%$ by number of the first generation of stars;
these massive stars would leave behind black holes with masses
$\mh\sim 10\msun$, 
which would rapidly fall to the cluster center and form 
black-hole binaries (\cite{pzm99}).
Subsequent close encounters will eject most
of the black holes, with perhaps a single binary remaining
at the cluster center after $\sim 10^9$ yr (\cite{pzm00}).
Binary black holes are also expected to form in 
galactic nuclei following galaxy mergers (\cite{bbr80});
here the characteristic masses are $\mh\sim 10^8\msun$.
The evolution of a binary supermassive black hole due to encounters 
with stars has been modelled by Makino (1997),
Quinlan \& Hernquist (1997) and others using $N$-body codes.
Quinlan \& Hernquist (1997) found that the wandering of the binary in
their simulations was a factor $5-10$ greater in
amplitude than expected on the basis of equation (\ref{equipart}),
and attributed this difference to inelastic scattering of
stars off the binary.

More speculatively, the massive halos of the Milky Way and other
galaxies may contain large numbers of black holes.
A population of stellar-mass, primordial black holes 
has been proposed to explain the gravitational microlensing results
toward the Large Magellanic Cloud (\cite{alc97}).
A small fraction of these black holes would be expected to 
form binaries through three-body interactions (\cite{nak97}). 
Alternatively, the halos of galaxies like the Milky Way
may consist of black holes of mass $\sim 10^6\msun$
(\cite{cba84}; \cite{lao85}) which would sink toward the galactic
center and form binary systems there (\cite{hur92}).

The contribution of inelastic scattering to the random motion of a 
massive binary can be estimated as follows.
Define $a$ to be the semi-major axis of the binary and $\m12=M_1+M_2$
its mass, with $\m12\gg\mf$.
The impact parameter $p$ corresponding to a closest-approach distance $r_p$ of a field star to the binary is
\beq
p^2 = r_p^2\left(1 + {2G\m12\over V^2 r_p}\right) 
\approx {2G\m12\over V^2} r_p
\label{grav}
\eeq
with $V$ the relative velocity at infinity.
The rate of encounters with stars that interact strongly with the binary, 
$r_p\lap a$, is then $\sim\nf (\pi p^2) \sf\approx 2\pi \nf G\m12 a/\sf$ with
$\nf$ the number density of field stars.
These stars are ejected with typical velocities $v_{ej}\approx \sqrt{G\m12/a}$, the binary orbital velocity (\cite{hif80}), 
yielding a rate of energy extraction from the binary of
\beq
\left|{d E_b\over dt}\right|\approx \left({1\over 2} \mf {G\m12\over a}\right) \times \left({2\pi\nf G\m12 a\over\sf}\right) \approx {\pi G^2\m12^2\rf\over\sf}
\label{eq1}
\eeq
with $\rf=\mf\nf$.
We can also write
\beq
\left|{d E_b\over dt}\right| = {d\over dt}\left({G(\m12/2)^2\over 2a}\right)
\label{eq2}
\eeq
(assuming $M_1=M_2$),
which, together with equation (\ref{eq1}), implies the well-known result that a hard binary hardens at a constant rate (\cite{heg75}):
\beq
{d\over dt}\left({1\over a}\right) \approx {8\pi G\rf\over\sf}.
\eeq
The hardening rate is usually expressed in terms of a dimensionless
constant $H$ as
\beq
{d\over dt}\left({1\over a}\right) \equiv H{G\rf\over\sf}.
\label{def_h}
\eeq
Three-body scattering experiments 
(\cite{hil83}; \cite{miv92}; \cite{qui96}) 
give $H\approx 15$ for a hard ($G\m12/a\gg\sf^2$), equal-mass, 
circular-orbit binary with $\m12\gg\mf$.

In terms of $H$, the energy extraction rate is given precisely by
\beq
\left|{d E_b\over dt}\right| = {d\over dt}\left({G\m12^2\over 8a}\right) = {G^2\m12^2\rf\over 8\sf} H.
\eeq
Momentum conservation implies that almost all of the binding energy released during an encounter will go into kinetic energy of the field star.
Thus
\beq
\delta E_b\approx \delta E_* \approx {1\over 2}\mf(\delta v_*)^2 \approx {1\over 2} \mf\left({\m12\over \mf}\delta v\right)^2
\eeq
with $\delta v$ the change in the center-of-mass velocity of the binary;
the last relation follows from conservation of momentum.
The rate of diffusion in velocity of the binary due to super-elastic scattering is then
\beq
\langle(\Delta v)^2\rangle_{S.E.} \approx {2\mf\over\m12^2} \left|{dE_b\over dt}\right| \approx 
{H\over 4}{G^2\mf\rf\over \sf}.
\eeq

This diffusion rate may be compared with Chandrasekhar's 
expression for a point mass initially at rest,
\beq
\langle (\Delta v)^2\rangle_{\cal C} = 
8(2\pi)^{1/2} {G^2 \mf\rf\over\sf}\ln\Lambda,
\eeq
with $\ln\Lambda\equiv\ln\left(p_{max}/p_{min}\right)$, the Coulomb logarithm (\cite{spi87}, eq. 2-12).
Both diffusion coefficients scale in the same way with the parameters
$(\mf,\rf,\sf)$ that define the field star distribution,
except for the weak variation implicit in the Coulomb logarithm.
The ratio of the two coefficients,
\beq
\sim{H\over 32\sqrt{2\pi}\ln\left(p_{max}/p_{min}\right)},
\label{ratio}
\eeq
therefore depends only on $\ln\Lambda$ and $H$, 
reflecting the fact that the Chandrasekhar coefficient is 
dominated by distant encounters while the effects of super-elastic 
scattering are limited to close encounters.
A more exact definition of $\ln\Lambda$, 
and appropriate choices for $p_{max}$ and $p_{min}$,
will be presented below; 
here we note that $p_{min}$ is of order $G\m12/\sf^2\equiv r_G$, 
the radius of gravitational influence of the massive binary.
If the distribution of field stars around the binary falls off steeply
over a radius comparable to $r_G$,
$p_{max}$ will be comparable to $p_{min}$ and $\ln\Lambda$ will be small.
Equation (\ref{ratio}) therefore predicts that super-elastic scattering
will increase the amplitude of the binary's random
velocity by a modest amount compared to 
the random velocity of a point particle with the same mass.

There is a second way in which energy exchange between field stars
and the binary will act to increase the magnitude of the
binary's random velocity.
Brownian motion represents a balance between dynamical friction
and scattering.
But field stars that interact strongly with the binary are ejected 
in nearly random directions, 
and this reduces the dynamical friction force that
they exert on the binary.
The velocity change experienced by a field star in a 
low-impact-parameter collision with a point-mass perturber 
is $\sim -2V$, corresponding to a $180^{\circ}$ change in its
direction (cf. equation \ref{dvp}).
When the point mass is replaced by a hard binary,
the field star is ejected in a nearly random direction and its mean
velocity change (averaged over many encounters with different
phases and orientations of the binary)
is therefore $\sim -V$ in a direction parallel to ${\bf V}$.
The drag force exerted on the massive object is proportional
to the mean velocity change of the field stars and hence the contribution
to the frictional force from close encounters is only $\sim 1/2$ as great
in the case of a binary as in the case of a point mass.

At first sight, this reduction in the frictional force might appear
to be significant. 
In the standard treatment (e.g. \cite{spi87}, eq. 2-46),
dynamical friction arises entirely from field stars with velocities 
less than that of the massive object.
If a massive binary were moving slowly,
all of these field stars would have small relative velocities as well 
and would interact strongly with it.
However it turns out (\S2.1) that the standard treatment is incorrect
in this regard and the reduction in dynamical friction is 
correspondingly more modest (\S4.1).

The approach adopted in the present paper consists of calculating the
dynamical friction and diffusion coefficients for a massive binary
under exactly the same approximations adopted by Chandrasekhar (1942)
in his analysis of gravitational encounters with a point mass.
In the case of a binary, this calculation requires the numerical 
integration of a large number of binary-field star interactions, i.e.
scattering experiments (\S 4).
The dynamical friction and diffusion coefficients can then be
computed by taking appropriate averages over the velocity changes
experienced by the field stars in these experiments.
Comparing the coefficients computed for the binary
to those computed by Chandrasekhar for a point mass gives the 
increase in the expected magnitude of the binary's random velocity
 compared to that of a point mass with $M=\m12$.
The restriction in the present paper to a massive binary, 
$\m12\gg\mf$, permits a number of simplications;
for instance, the mass of the field star can be neglected in the
scattering experiments, and the values of the dynamical friction and
diffusion coefficients need only be computed in the limit $v\ll\sf$ 
appropriate to a massive body that is near energy equipartition with
much lighter stars.

It became clear after this investigation was begun that the Chandrasekhar
coefficients as they are usually presented could not be directly
compared with the results of the scattering experiments.
Starting with Chandraskehar (1943a), the dependence of the Coulomb
logarithm on the velocity of the field star has generally been ignored
when integrating over the field star velocity distribution
(e.g. \cite{rmj57}; \cite{hen73}).
This approximation is valid in the case of a test star whose velocity
is comparable to that of the field stars but not when $v\ll\sf$, 
which is the case of interest here.
Thus in \S 2 the Chandrasekhar coefficients are re-derived without
the usual approximations.
One of the interesting results of this re-derivation is that the
dynamical friction force on a slowly-moving body is found to come 
primarily from field stars with $\vf>v$.
In the standard treatment, all of the frictional force comes from
field stars moving more {\it slowly} than the test star.

An important preliminary step is to relate the equilibrium velocity 
distribution of a test mass (single star or binary) to its dynamical 
friction and diffusion coefficients, via the Fokker-Planck equation.
This is done in \S 2.
Applications of the results to various physical systems are discussed
in \S 5.

\section{Fokker-Planck Equation}

We wish to compute the equilibrium velocity distribution $f(v)$ of a 
massive binary that interacts via gravitational encounters with 
individual stars;
${\bf v}$ is the velocity of the binary's center of mass.
We assume that the time scale over which this equilibrium is reached is 
much shorter than the time scale over which the orbital elements of the 
binary change; this assumption is discussed in more detail in \S 5.2.
We also ignore a possible dependence of $f$ on the direction of ${\bf v}$, appropriate if the orientation of the binary relative to ${\bf v}$ is rapidly changing and if the field stars are distributed isotropically in velocity.
The dependence of the dynamical friction force on the orientation of the binary is discussed briefly in \S4.1.
Because of the assumed large mass ratio, $\m12\gg\mf$, even close encounters will produce only small deflections in the trajectory of the binary.
Hence we may describe the evolution of $f$ via the Fokker-Planck equation (\cite{cha43b}).
In a steady state, this equation implies
\beq
0=-N(v)\langle\Delta v\rangle + {1\over 2}{\partial\over\partial v}\left[N(v)\langle(\Delta v)^2\rangle\right]
\label{fp1}
\eeq
where
\beq
4\pi v^2 f(v) dv = N(v) dv .
\label{defn}
\eeq
The coefficients $\langle\Delta v\rangle$ and $\langle(\Delta v)^2\rangle$ 
are defined in the usual way as sums, over a unit interval of time,
of $\Delta v$ and $(\Delta v)^2$ due to encounters with field stars.

Define $\Delta\vp$ and $\Delta\vt$ to be changes in $v$ in directions parallel and perpendicular to ${\bf v}$.
We have
\beq
\Delta v = \left[\left(v+\Delta\vp\right)^2 + (\Delta\vt)^2\right]^{1/2} - v.
\eeq
Expanding to second order in the small quantities $\Delta\vp$ and $\Delta\vt$ and taking means,
the coefficients in equation (\ref{fp1}) may be expressed in terms of the
usual dynamical friction and diffusion coefficients as
\beq
\langle\Delta v\rangle = \df + {1\over 2}{\langle(\Delta\vt)^2\rangle\over v},\ \ \ \  \langle(\Delta v)^2\rangle = \langle(\Delta\vp)^2\rangle.
\label{dfv}
\eeq
Substituting equations (\ref{defn}) and (\ref{dfv}) into equation (\ref{fp1}) yields\footnote{
In equation (\ref{fp2}) and below, 
the parentheses in expressions like $\langle(\Delta\vp)^2\rangle$ are dropped.}
\beq
0=f\left[\df + {1\over 2v}\left(\dvt - 2\dvp\right)\right] - {1\over 2}{\partial\over\partial v}\left(f\dvp\right).
\label{fp2}
\eeq
Equation (\ref{fp2}) may also be derived from equation (2-71) of Spitzer 
(1987) after replacing $\partial E$ in that equation by $v\partial v$
and expressing $\langle\Delta E\rangle$ and $\langle\Delta E^2\rangle$ in terms of $\df$, $\dvp$ and $\dvt$.

In a steady state, we expect that $v$ will be of order
$\sqrt{\mf/\m12}\ll 1$ times the velocity dispersion $\sf$ of the field stars.
Hence it is appropriate to expand the dynamical friction and diffusion coefficients about $v=0$:
\begin{mathletters}
\begin{eqnarray}
\df &=& -Av+ Bv^3\ldots , \label{expand1} \\
\dvp &=& C + Dv^2\ldots , \label{expand2} \\
\dvt &=& 2(E + Fv^2)\ldots\label{expand3}
\label{123}
\end{eqnarray}
\end{mathletters}
Substituting these expressions into equation (\ref{fp2}) gives
\beq
0 = \left(-Av^2 + Bv^4 +E+Fv^2 -C - 2Dv^2 \right)f - {v\over 2}\left(C+Dv^2\right){\partial f\over\partial v}.
\label{eq0}
\eeq
Setting $v=0$ gives $C=E$, i.e. the diffusion rate of a stationary particle must be independent of direction.
Cancelling additional terms, we find to lowest order in $v$
\beq
0=\left(A+2D-F\right)f + {C\over 2}{\partial f\over\partial v}{1\over v}.
\eeq

The standard Chandrasekhar coefficients for a point particle of mass $M$ interacting with stars of mass $\mf$ (e.g. Spitzer 1987, eqs. 2-52 -- 2-54) give, after expanding about $v=0$:
\begin{mathletters}
\begin{eqnarray}
\ac &=& {4\sqrt{2\pi}\over 3}{G^2\mf^2\nf\ln\Lambda\over\sigma_f^3}
\left(1+{M\over\mf}\right), 
\ \ \ \ \ \ \bc = {3\over 10} {\ac\over\sf^2}, \label{ac} \\
\cc &=& {8\sqrt{2\pi}\over 3}{G^2\mf^2\nf\ln\Lambda\over\sigma_f}, 
\ \ \ \ \ \ \dc = -{3\over 10}{\cc\over\sf^2}, \label{cc} \\
\ec &=& \cc,\ \ \ \ \ \ \fc = -{1\over 10}{\cc\over\sf^2}.\label{ec}
\end{eqnarray}
\end{mathletters}
Thus
\begin{mathletters}
\begin{eqnarray}
\ac+2\dc-\fc &=& \ac-{\cc\over 2\sf^2} \\
&=& -{4\sqrt{2\pi}\over 3}{G^2M\mf\nf\ln\Lambda\over\sigma_f^3} \\
&\approx& \ac
\end{eqnarray}
\end{mathletters}
where the last relation is valid when $M\gg\mf$.
In the case of the more general coefficients describing the 
motion of a massive binary, 
$A$ will still be of order $\m12/\mf$ times $D$ and $F$ 
and we may likewise approximate $(A+2D-F)$ by $A$. 
Thus
\beq
0\approx Af + {C\over 2}{\partial f\over\partial v}{1\over v}
\eeq
which has solution
\begin{mathletters}
\begin{eqnarray}
f(v) & = & f_0 e^{-v^2/2\sigma^2},\label{soln1} \\
\sigma^2 & = & {C\over 2A}.\label{soln2}
\end{eqnarray}
\end{mathletters}

Substituting $\ac$ for $A$ and $\cc$ for $C$ gives the well-known result for the equipartition velocity dispersion of a point mass,
\beq
\sigma_{\cal C} = \left({\mf\over M}\right)^{1/2}\sf.
\label{equipart2}
\eeq
It will be useful in what follows to express the coefficients $A$ and $C$ for a massive binary in terms of their values for a point mass with $M=\m12$, i.e. in terms of the ratios
\beq
R_1 \equiv {A\over \ac},\ \ \ \ R_2 \equiv {C\over \cc}.
\label{ratios}
\eeq
(The Chandrasekhar coefficients $\ac$ and $\cc$ will be defined 
slightly differently below.)
The velocity dispersion of the binary is then
\beq
\sigma = \left({R_2\over R_1}\right)^{1/2}\left({\mf\over\m12}\right)^{1/2}\sigma_f.
\label{eq_equi}
\eeq
Thus the problem of determining the equilibrium velocity distribution of the binary's center-of-mass motion is reduced to determining the ratios of $\df$ and $\dvp$, the dynamical friction and diffusion coefficients for the binary, 
to the Chandrasekhar coefficients for a point mass in the low-$v$ limit.
We note that $R_2$ can equally well be defined as $(C+2E)/(\cc+2\ec)$,
i.e. in terms of the diffusion coefficient $\langle\Delta v^2\rangle=\dvp+\dvt$, in the limit $v\rightarrow 0$. 

\section{Chandrasekhar Coefficients for a Point Mass}

The Chandrasekhar coefficients for a point mass in the form 
given above (eqs. \ref{ac}--\ref{ec}) embody two approximations: the argument of $\ln\Lambda$ was assumed constant when integrating over the distribution of field-star velocities; and the non-dominant terms, which are of order $1/\ln\Lambda$ or $1/\Lambda$ times the dominant terms, were ignored.
Both approximations are reasonable when the velocity of the test star is not too different from the typical velocity of the field stars, and when the integration over impact parameters extends to distances much greater than $GM/\sf^2$.
However it is impossible to reproduce these approximations precisely when carrying out the scattering experiments discussed below; 
hence in order to compute the ratios (eqs. \ref{ratios}) between the coefficients for the binary and for a point mass, we must first rederive the Chandrasekhar coefficients without the usual approximations.
Of particular interest here is the velocity dependence of the Coulomb logarithm.
Chandrasekhar (1943a) and White (1949) noted that this dependence implies that some of the dynamical friction force comes from field stars moving more rapidly than the test mass; 
in the standard treatment, 
where $\ln\Lambda$ is removed from the integral over field star velocities,
all of the friction is produced by field stars with $\vf<v$.
An exact analysis in the case of interest here, $v\ll\sf$, has apparently never been carried out.
We will find in fact that almost all of the frictional force comes from field stars with $\vf>v$ in the low-velocity limit.
We also derive a precise expression for the Coulomb logarithm in this limit.

\subsection{Dynamical Friction}

The velocity change of a test particle of mass $M$ in one encounter with a field particle of mass $\mf\ll M$ is (\cite{spi87}, eq. 2-19)
\beq
\Delta \vp = -2V{m_f\over M}{1\over 1+p^2/p_0^2}
\label{dvp}
\eeq
where $V$ is the relative velocity at infinity, $p$ is the impact parameter, and $p_0\equiv GM/V^2$.
Multiplying by $2\pi pn_fVdp$, with $n_f$ the number density of field stars, and integrating over $p$ gives
\beq
\overline{(\Delta\vp)} = -{2\pi G^2M\mf n_f\over V^2}\ln\left(1+p^2_{max}/p_0^2\right),
\label{coef1}
\eeq
the rate of change of $\vp$ due to encounters with field stars of velocity $V$.

The dynamical friction coefficient is (Spitzer 1987, eq. 2-28)
\begin{mathletters}
\begin{eqnarray}
\df & = & \int \ff({\bf v}_f)\, \overline{(\Delta\vp)}\,{v-{v_f}_x\over V}\bf{dv}_f \label{df} \\
& = & -2\pi G^2M m_fn_f\int \ff({\bf v}_f) {v-v_{f,x}\over V^3} \ln\left(1+{p_{max}^2 V^4\over G^2 M^2}\right)\bf{dv}_f,
\end{eqnarray}
\end{mathletters}
with $v$ the velocity of the (massive) test particle, 
$\ff$ the distribution of field-star velocities (normalized to unit total number) and ${\bf V} = {\bf v} - {\bf v}_f$; 
${\bf v}$ has been assumed parallel to the $x$-axis.
Following Chandrasekhar (1943a), 
we can represent the velocity-space volume element in terms of $v_f$ and $V$; 
then writing
\beq
v-{v_f}_x = {V^2+v^2-v_f^2\over 2v}
\eeq
gives
\beq
\df = 
-2\pi^2 G^2M m_fn_fv^{-2}\int_0^{\infty} dv_f\, v_f\, \ff(v_f) \int_{|v-v_f|}^{v+v_f} dV\left(1+{v^2-v_f^2\over V^2}\right)\ln\left(1+{p^2_{max}V^4\over G^2M^2}\right)
\label{df2}
\eeq
where $\ff$ is henceforth assumed isotropic.

In the standard approximation (e.g. \cite{rmj57}), 
the velocity dependence of the logarithmic term is ignored, and one writes
\beq
\ln\left(1+{p^2_{max}V^4\over G^2M^2}\right) \approx 2\ln\Lambda, \ \ \ \ \ \ \Lambda \equiv {p_{max}\over p_{min}}.  
\eeq
The dynamical friction coefficient then becomes
\beq
\df = -4\pi(4\pi G^2 Mm_f n_f) \int_0^{\infty}dv_f \left({v_f\over v}\right)^2 \ff(v_f) H_1(v,\vf,p_{max}) 
\label{dfh1}
\eeq
with
\beq
H_1 = \left\{ \begin{array}{ll}
	\ln\Lambda & \mbox{if $v>v_f$,} \\
	0 & \mbox{if $v<v_f$}.
	\end{array}
	\right.
	\label{twoh}
\eeq
Neither $p_{max}$ nor $p_{min}$ are well-defined;
$p_{min}$ is typically set to some multiple of $GM/\sf^2$
(e.g. Spitzer 1987, eq. 2-14).
Equation (\ref{twoh}) reproduces the well-known result that only field stars with $v_f<v$ contribute to the frictional force.

When the velocity of the test particle is sufficiently low,
as in the case of a massive particle near equipartition with 
lighter field particles, 
the logarithm in equation (\ref{df2}) will be 
close to zero for {\it all} stars with $v_f<v$.
Hence it is unclear whether removing the logarithm from the integrand is a reasonable approximation, or what the proper definition of $\ln\Lambda$ (particularly $p_{min}$) in the final expression should be.
The frictional force in the low-$v$ limit must either be much less than implied by equation (\ref{dfh1}),
or else most of the frictional force must come from field stars with $v_f>v$.

In fact the latter is the case, as we now show.
Returning the logarithmic term to the integrand in equation (\ref{dfh1})
gives
\beq
H_1(v,v_f,p_{max}) = {1\over 8v_f}\int_{|v-v_f|}^{v+v_f} dV\left(1 + {v^2-v_f^2\over V^2}\right) \ln\left(1 + {p^2_{max}V^4\over G^2 M^2}\right).
\label{hofv}
\eeq
This function is plotted in Figure 1a.
For low test-star velocities, $v\lap\sqrt{GM/p_{max}}$,
$H_1$ deviates strongly from a step function;
it reaches a peak at $\vf\approx \sqrt{GM/p_{max}}$ and its peak amplitude varies as $\sim v^3$ due to the velocity dependence of the logarithm.
Figure 1b plots the (normalized) integrand of the dynamical friction integral, 
equation (\ref{dfh1}), in the limits of low and high $v/\sf$.
The field-star velocity distribution was assumed to be a Maxwellian:
\beq
\ff(v_f) = {1\over (2\pi\sf^2)^{3/2}} e^{-v_f^2/2\sigma_f^2}.
\label{max}
\eeq
For low $v/\sf$, the integrand of equation (\ref{dfh1}) may be shown to vary approximately as
\beq
{\xf^3 e^{-\xf^2/2}\over 1+R^2\xf^4}
\label{thing}
\eeq
where
\beq
\xf \equiv {\vf\over \sigma_f}, \ \ \ \ 
R \equiv {p_{max}\over p_f}, \ \ \ \ 
p_f \equiv {GM\over\sf^2}.
\eeq
This function peaks at $\xf\approx R^{-1/2}$.
In the case of high $v/\sf$, on the other hand, the integrand varies as
\beq
\xf^2 e^{-\xf^2/2}
\label{thing2}
\eeq
which peaks at $\xf=1$.
Thus for $R\equiv p_{max}/\pf\approx 1$, 
the dynamical friction force comes from field stars with $\vf\sim\sf$, 
whether the test star is moving much faster or much slower than the typical field star!
For $R\gap1$, the frictional force in the low-$v$ limit comes from field stars with $\vf\lap \sf/R^{1/2}$;
thus for any $v\lap \sf/R^{1/2}$, 
i.e. for $M\gap R\mf$,
most of the stars producing the frictional force will be moving faster than the test star.  
The appropriate choice for $R$ is discussed in $\S5$ where it is argued that $R\gap 1$ in many cases of interest.

The complete expression for the dynamical friction coefficient is
\begin{mathletters}
\begin{eqnarray}
\df & = & -16\sqrt{\pi}G m_f p_f n_f F(v/\sf,p_{max}/\pf), \label {coef0} \\
F(x,R) & = & x^{-2}\int_0^{\infty} dy\: y^2 e^{-y^2} H_1(x,\sqrt{2}y,R), 
\label{coef2} \\
H_1(x,z,R) & = & {1\over 8z} \int_{|x-z|}^{x+z} dw \left(1 + {x^2-z^2\over w^2}\right)\ln\left(1+R^2w^4\right) \label {coef3}
\end{eqnarray}
\end{mathletters}
where $x\equiv v/\sf$ and $R$ and $\pf$ are defined above.
The function $F(v,p_{max})$ is plotted in Figure 2.
The dependence of $\df$ on $v$ is seen to be a function of 
$R$. 
By contrast, when $\ln\Lambda$ is assumed fixed, 
the $v$-dependence of $\df$ is independent of $p_{max}/p_{min}$ (Spitzer 1987, eq. 2-52; Fig. 2).

The low-$v$ limit of the frictional force may be derived 
by careful manipulation of equations (\ref{coef0} - \ref{coef3}).
A more transparent route, useful in what follows, 
is to first change variables in equation (\ref{df}).
\footnote{I am grateful to M. Milosavljevi\' c for pointing this out.}
Representing the velocity-space volume element in terms of $V$ and $\lambda$, where $\lambda$ is the direction cosine between ${\bf V}$ and ${\bf v}$
\beq
\lambda = {(v\hat{\bf e}_x - {\bf v}_f)\cdot v\hat{\bf e}_x\over Vv} = {v-v_{f,x}\over V},
\eeq
gives
\begin{mathletters}
\begin{eqnarray}
\df & = & -2\pi G^2M \mf\nf \times 2\pi\int_0^{\infty} dV V^2 \int_{-1}^1 d\lambda {\lambda\over V^2} \ff(v_f)\ln\left(1+{p^2_{max}V^4\over G^2M^2}\right) \\
& = & -4\pi^2 G^2M m_fn_f\int_0^{\infty} dV\ln\left(1+{p^2_{max}V^4\over G^2M^2}\right)\int_{-1}^1 d\lambda\lambda \ff(v_f).
\end{eqnarray}
\end{mathletters}
Again substituting equation (\ref{max}) for $\ff$ and integrating over $\lambda$,
\begin{eqnarray}
\df & = & -{2\sqrt{2\pi} G^2M\mf\nf\over\sf^3} e^{-v^2/2\sf^2} \int_0^{\infty} dV\: e^{-V^2/2\sf^2} \ln\left(1+{p^2V^4\over G^2M^2}\right) \nonumber \\
&\times&{\sf^4\over v^2V^2} \left[{vV\over\sf^2}\cosh\left({vV\over\sf^2}\right) - \sinh\left({vV\over\sf^2}\right)\right].
\end{eqnarray}
Taking the limit $v\ll\sf$, and writing by analogy with equation 
(\ref{expand1})
\beq
\df = -\ac v + \bc v^3 \dots ,
\eeq
we find
\begin{mathletters}
\begin{eqnarray}
\ac& = & {2\sqrt{2\pi}\over 3} {G^2M m_fn_f\over\sigma_f^3}\int_0^{\infty} dz\ e^{-z} \ln\left(1+4 R^2z^2\right) , \\
\bc& = & {\sqrt{2\pi}\over 15} {G^2M\mf\nf\over\sigma_f^5}\int_0^{\infty} dz\ e^{-z} (5-2z)\ln\left(1+4 R^2z^2\right).
\end{eqnarray}
\end{mathletters}
Thus Hooke's law ($\df\propto -v$) is recovered for $v\ll\sf$.

Of primary interest here is the leading coefficient, which may be written
\begin{mathletters}
\begin{eqnarray}
\ac& = & {4\sqrt{2\pi}\over 3} {G^2M m_fn_f\over\sigma_f^3} G(R), \label{df0} \label{acnew} \\ 
G(R) & \equiv & {1\over 2} \int_0^{\infty} dz\ e^{-z} \ln\left(1+4 R^2z^2\right).
\label{gofr}
\end{eqnarray}
\end{mathletters}
A reasonable approximation to $G(R)$, valid for $R\gap 1$, is
\beq
G(R)\approx \ln\sqrt{1+2R^2} \approx \ln\sqrt{1+{2p^2_{max}\sf^4\over G^2M^2}}
\label{gofr2}
\eeq
(Figure 3).
Comparing equations (\ref{df0}) and (\ref{gofr2}) with the standard 
dynamical friction coefficient for a point mass in the 
low-test-particle-velocity limit,
equation (\ref{ac}), we find
\beq
\ln\Lambda\equiv\ln\sqrt{1+{p^2_{max}\over p^2_{min}}} \approx 
\ln\sqrt{1 + 2R^2} = \ln\sqrt{1 + {2p^2_{max}\sf^4\over G^2M^2}}
\label{approx}
\eeq
or
\beq
p_{min}\approx {GM\over \sqrt{2}\sf^2}.
\label{pmin}
\eeq
Spitzer (1987, eq. 2-14) advocates
\beq
p_{min} \approx {G(\mf+M)\over 6\sf^2}
\eeq
in the case that the test and field stars have similar masses
and velocities.
When $M\gg\mf$, the denominator in Spitzer's expression
should be reduced by a factor $\sim 2$ to account for the lower relative velocities, giving a value for $p_{min}$ that is not too different from the value derived here.
This agreement is not fortuitous: it is a result of the fact
that the field stars responsible for producing the
frictional force have roughly the same velocity distribution
whether the test star is moving rapidly or slowly.

\subsection{Diffusion}

The squared velocity changes of the test particle in one encounter  
with a field star are (cf. Spitzer 1987, equations 2-18 and 2-19):
\beq
(\Delta \vp)^2 = 4V^2{m_f^2\over M^2}{1\over (1+p^2/p_0^2)^2},\ \ \ \ 
(\Delta \vt)^2 = 4V^2{m_f^2\over M^2}{p^2/p_0^2 \over (1+p^2/p_0^2)^2}.
\label{scat1}
\eeq
Multiplying by $2\pi pn_fVdp$ and integrating over $p$ as before,
\begin{mathletters}
\begin{eqnarray}
\overline{(\Delta\vp)^2} & = & {4\pi G^2 n_fm_f^2\over V}\left({p^2_{max}/p_0^2\over 1+p_{max}^2/p_0^2}\right), \label{scat2} \\
\overline{(\Delta\vt)^2} & = & {4\pi G^2 n_fm_f^2\over V}\left[\ln\left(1+{p_{max}^2\over p_0^2}\right) - {p^2_{max}/p_0^2\over 1+p_{max}^2/p_0^2}\right],
\label{scat3}
\end{eqnarray}
\end{mathletters}
the rates of change due to encounters with field stars whose velocity at infinity is $V$.

The expressions just given refer to velocity changes with respect to the direction of the initial relative velocity vector ${\bf V}$.
We need to transform to a frame in which the test particle has velocity ${\bf v}$.
Following Spitzer (1987, eqs. 2-24, 2-25), but retaining the contributions from the non-dominant terms, gives in the fixed frame
\begin{mathletters}
\begin{eqnarray}
\langle\Delta v_i\Delta v_j\rangle & = & (\hat{\bf e}_i\cdot\hat{\bf e}_1')(\hat{\bf e}_j\cdot\hat{\bf e}_1')\overline{(\Delta\vp)^2} + {1\over 2}\left[(\hat{\bf e}_i\cdot\hat{\bf e}_2')(\hat{\bf e}_j\cdot\hat{\bf e}_2') + (\hat{\bf e}_i\cdot\hat{\bf e}_3')(\hat{\bf e}_j\cdot\hat{\bf e}_3')\right]\overline{(\Delta\vt)^2} \\
& = & {V_iV_j\over V^2}\left[\overline{(\Delta\vp)^2} - {1\over 2}\overline{(\Delta\vt)^2}\right] + {1\over 2}\delta_{ij}\overline{(\Delta\vt)^2}
\end{eqnarray}
\end{mathletters}
or
\begin{mathletters}
\begin{eqnarray}
\langle\Delta v_1\Delta v_1\rangle & = & {V_x^2\over V^2}\overline{(\Delta\vp)^2} + {1\over 2}\left(1-{V_x^2\over V^2}\right)\overline{(\Delta\vt)^2},\\
\langle\Delta v_2\Delta v_2\rangle + \langle\Delta v_3\Delta v_3\rangle &
 = & {V_y^2+V_z^2\over V^2}\overline{(\Delta\vp)^2} + \left(1-{1\over 2}{V_y^2+V_z^2\over V^2}\right)\overline{(\Delta\vt)^2}
\end{eqnarray}
\end{mathletters}
where the ${\bf e}_1$ axis is oriented parallel to ${\bf v}$.
The final integrations over field-star velocities are then
\begin{mathletters}
\begin{eqnarray}
\dvp & = & {2\pi\over v}\int_0^{\infty} dv_f\ v_f \ff({\bf v}_f)\int_{|v-v_f|}^{v+v_f} dV\ V\ \langle\Delta v_1\Delta v_1\rangle, 
\label{int3a} \\
\dvt & = & {2\pi\over v}\int_0^{\infty} dv_f\ v_f \ff({\bf v}_f)\int_{|v-v_f|}^{v+v_f} dV\ V\left[ 
\langle\Delta v_2\Delta v_2\rangle + \langle\Delta v_3\Delta v_3\rangle \right].
\label{int3b}
\end{eqnarray}
\end{mathletters}
Writing 
\beq
V_x = {V^2+v^2-v_f^2\over 2v},\ \ \ \ \ \ \ V_y^2+V_z^2 = 1 - V_x^2,
\eeq
we find as before
\begin{mathletters}
\begin{eqnarray}
\dvp & = & {8\pi\over 3}(4\pi G^2m_f^2n_f)v \int_0^{\infty}dv_f \left({v_f\over v}\right)^2 \ff(v_f) H_2(v,v_f,p_{max}), \\
\dvt & = & {8\pi\over 3}(4\pi G^2m_f^2n_f)v \int_0^{\infty}dv_f \left({v_f\over v}\right)^2 \ff(v_f) H_3(v,v_f,p_{max}), \\
H_2(v,v_f,p_{max}) & = &{3\over 8v_f}\int_{|v-v_f|}^{v+v_f} dV \left[1-{V^2\over 4v^2}\left(1 + {v^2-v_f^2\over V^2}\right)^2\right]\ln\left(1 + {p^2_{max}V^4\over G^2 M^2}\right) \\ 
& + & \left[{3\over 4}{V^2\over v^2}\left(1+{v^2-v_f^2\over v^2}\right)-1\right]{p^2_{max}V^4/G^2M^2\over 1+p^2_{max}V^4/G^2M^2} ,\nonumber \\
H_3(v,v_f,p_{max}) & = &{3\over 8v_f}\int_{|v-v_f|}^{v+v_f} dV \left[1+{V^2\over 4v^2}\left(1 + {v^2-v_f^2\over V^2}\right)^2\right]\ln\left(1 + {p^2_{max}V^4\over G^2 M^2}\right) \nonumber \\
& + & \left[1-{3\over 4}{V^2\over v^2}\left(1+{v^2-v_f^2\over v^2}\right)\right]{p^2_{max}V^4/G^2M^2\over 1+p^2_{max}V^4/G^2M^2}.
\end{eqnarray}
\end{mathletters}

In the standard approximation, 
$\ln\Lambda$ is assumed to be a constant and the non-dominant terms are ignored, yielding
\beq
H_2 = \left\{ \begin{array}{ll}
	\ln\Lambda\left({v_f\over v}\right)^2 & \mbox{if $v>v_f$,} \\
	\ln\Lambda\left({v\over v_f}\right)  & \mbox{if $v<v_f$.}
	\end{array}
	\right.
\eeq
\beq
H_3 = \left\{ \begin{array}{ll}
	\ln\Lambda\left(3 - {v_f^2\over v^2}\right) & \mbox{if $v>v_f$,} \\
	2\ln\Lambda\left({v\over v_f}\right)  & \mbox{if $v<v_f$.}
	\end{array}
	\right.
\eeq

We are again primarily interested in the low-$v$ limits.
Returning to equations (\ref{int3a}) and (\ref{int3b}),
expressing the integral in the same $(V,\lambda)$ variables as in $\S3.1$, 
and expanding to second order in $v/\sigma_f$ gives
\begin{mathletters}
\begin{eqnarray}
\dvp &=& \cc + \dc v^2\ldots , \\
\dvt &=& 2(\ec + \fc v^2)\ldots,
\end{eqnarray}
\end{mathletters}
\begin{mathletters}
\begin{eqnarray}
\cc & = & {8\over 3}\sqrt{2\pi}{G^2m_f^2n_f\over\sf}\int_0^{\infty} dz\ e^{-z}\ln\left(1+4R^2z^2\right), \\
\dc & = & -{2\over 15}\sqrt{2\pi}{G^2m_f^2n_f\over\sf^3}\int_0^{\infty} dz\ e^{-z}\left(5+2z-2z^2\right)\ln\left(1+4R^2z^2\right), \\
\ec & = & \cc ,  \\
\fc & = & -{2\over 15}\sqrt{2\pi}{G^2m_f^2n_f\over\sf^3}\int_0^{\infty} dz\ e^{-z}\left(5-6z+z^2\right)\ln\left(1+4R^2z^2\right).  
\end{eqnarray}
\end{mathletters}
The leading terms are
\beq
\cc=\ec= {8\sqrt{2\pi}\over 3} {G^2\mf^2\nf\over\sf} G(R),
\label{cande}
\eeq
with $G(R)$ again given by equation (\ref{gofr}).
Comparing equation (\ref{cande}) with equation (\ref{df0}),
we see that the ratios between the three diffusion coefficients at low $v$ are exactly the same as in the standard treatment with $\ln\Lambda$ assumed constant.
Thus we predict the same equilibrium velocity dispersion for a point mass
as in equation (\ref{equipart2}), 
$\sigma^2=\cc/2\ac=(\mf/M)\sf^2$.

In what follows, the expressions (\ref{acnew}) and (\ref{cande}) will be used
to define $\ac$ and $\cc$ when computing the ratios $R_1$ and $R_2$ of the
dynamical friction and diffusion coefficients of the binary to those
of a point mass (eqs. \ref{ratios}).

\section{Scattering Experiments}

The dynamical friction and diffusion coefficients for a massive binary were computed by carrying out a large number of scattering experiments, in the manner of Hills (1983) and Quinlan (1996).
In the limit of large mass ratio $\m12/\mf$, 
the field star may be treated as a massless particle moving in the potential of the two black holes (\cite{miv92}), 
which follow unperturbed Keplerian trajectories about their fixed center of mass with semimajor axis $a$.
From the changes in the field star's velocity during the encounter we can infer the corresponding changes in the binary's center-of-mass motion.
Let the initial velocity of the field particle be ${\bf \vf}_0 = {\vf}_0\hat{{\bf e}_x}$ and its final velocity $({\vf'}_x\hat{\bf e}_x + 
{\vf'}_y\hat{\bf e}_y + {\vf'}_z\hat{\bf e}_z)$.
The velocity changes $(\delta v_x, \delta v_y, \delta v_z)$ 
of the binary's center of mass following a single collision are
\begin{mathletters}
\begin{eqnarray}
\delta v_x & = & -{\mf\over\m12}\left({\vf}_x' - {\vf}_0\right) = -{\mf\over\m12}\delta{\vf}_x, \\
\delta v_y & = & -{\mf\over\m12} {\vf}_y' = -{\mf\over\m12}\delta{\vf}_y, \\
\delta v_z & = & -{\mf\over\m12} {\vf}_z' = -{\mf\over\m12}\delta{\vf}_z.
\end{eqnarray}
\end{mathletters}
The dynamical friction and diffusion coefficients for the binary
are obtained by taking the appropriate averages, over an ensemble of trajectories, of the $\delta v_i$'s and their squares, as discussed in more detail below.

Orbits were integrated using the routine D0P853 of Hairer, N\o rsett \& Wanner (1991), an 8(6)th order embedded Runge-Kutta integrator.
The routine automatically adjusts the integration time step to keep the fractional error per step below some level $TOL$, which was set to $10^{-9}$.
The forces from the black holes were not softened.
Each field star was assumed to begin at a position $(x,y,z)=(\infty,p,0)$ and was advanced from $r=\infty$ to $r=50a$ along a Keplerian orbit about a point mass $\m12$.
The integrations were terminated when the star had moved a distance from the binary that was at least $100$ times its initial distance with positive energy, or when the number of integration steps exceeded $10^6$.
The latter condition was met only for the integrations with the lowest initial velocities, and even then, for much fewer than $1\%$ of the integrations.
These stars were on orbits that were weakly bound to the binary and that needed to make many revolutions before being expelled.
The incomplete integrations were not included when computing the diffusion coefficients below.

The binary was given unit total mass, $M_1=M_2=1/2$ and its center of mass was fixed at the origin.
The binary's orbit was assumed to be circular in all of the integrations described below.
The orientation of the binary's orbital plane, 
and its initial phase, were chosen randomly for each integration.
The angle $\xi$ between the normal to the binary orbital plane and the 
initial relative velocity was stored for later use,
since velocity changes in the field star are expected to 
depend systematically on $\xi$.
However most of the results presented below are based on averages over 
$\xi$ and over the phase of the binary, 
appropriate if the orientation of the binary relative to ${\bf v}$ is changing rapidly due to encounters.

Initial velocities for the scattering experiments were assigned one of the discrete values $K\times \vb = K\times\sqrt{G\m12/a}$ with $K=(0.01,0.02,0.03,0.05,0.1,0.2,0.3,0.5,1,2,3,5,10)$.
$10^6$ orbits were integrated for each value of $K$, for a total of $1.3\times 10^7$ integrations.
The impact parameters $p$ for each $V$ were chosen from one of $20$ intervals corresponding to ranges in scaled pericenter distance $r_p/a$ of 
$[0,0.001], [0.001,0.003], [0.003,0.01]$, 
$[0.01,0.02], [0.02,0.03], [0.03,0.05]$,
$[0.05,0.1], [0.1,0.2], [0.2,0.3]$, 
$[0.3,0.5], [0.5,1], [1,2]$, 
$[2,3], [3,5], [5,10]$, 
$[10,20], [20,30], [30,50]$, $[50,100]$ and $[100,200]$.
Impact parameters were chosen randomly and uniformly in $p^2$ within each interval.
Unless otherwise indicated, distances and velocities given below are in program units of $a$ and $(G\m12/a)^{1/2}$ respectively.

A number of checks of the numerical integrations were carried out.
The field-star velocity changes were used to recompute the binary 
hardening rates calculated by Mikkola \& Valtonen (1992), 
Hills (1983,1992) and Quinlan (1996).
The agreement was excellent.
The relation:
\beq
H_1(V) = {4v\over G^2M^2\nf}\left[2V\overline{(\Delta\vp)} + \left( 
\overline{(\Delta\vp)^2} + \overline{(\Delta\vt)^2}\right)\right],
\eeq 
was also checked;
here $H_1$ is the velocity-dependent hardening rate defined by Quinlan (1996),
and the diffusion coefficients on the right hand side refer to the field
stars.
A weaker check (since it does not depend on the details of binary-field star
interactions)
consisted of verifying that the diffusion coefficients 
$\df$ and $\dvt$
for the binary tended to those of a point mass for large $p_{max}$.

\subsection{Dynamical Friction}

For a binary initially at rest, the acceleration of its center of mass due to encounters with field stars of initial velocity $V$ is
\beq
\langle\delta v_x(V,p_{max})\rangle = 
2\pi\nf V \int_0^{p_{max}} dp\ p\ \overline{\delta{v}}_x(V,p) =
-2\pi\nf\left({\mf\over\m12}\right) V \int_0^{p_{max}} dp\ p\ 
\overline{{\delta{\vf}_x}}(V,p)
\eeq
where $\overline{{\delta{\vf}_x}}(V,p)$ is the mean change in the $x$-component of the field star velocity, averaged over many encounters with the binary at fixed $(V,p)$.
We are interested in the ratio between the numerically-computed coefficients and the exact expression for a point mass (equation \ref{coef1}).
This ratio is
\begin{mathletters}
\begin{eqnarray}
D_1(V,p_{max}) & = & -{V^3\over \ln\left(1+p^2_{max}/p_0^2\right)G^2\m12^2}\int_0^{p_{max}} dp\ p\ {\overline{\delta {\vf}}_x} \\ 
& = & -{1\over 2\ln\left(1+p^2_{max}/p_0^2\right)} \left({V\over\vb}\right)^3 \int_0^{p_{max}/a} d\left({p\over a}\right)^2 \left({{\overline{\delta v}}_x\over \vb}\right)
\label{d1}
\end{eqnarray}
\end{mathletters}
with $\vb^2\equiv G\m12/a$ and $p_0\equiv G\m12/V^2$.

Figure 4 shows values of $D_1$, averaged over all orientations of the binary, for several values of $V$.
Generally $D_1<1$ due to randomization by the binary of the velocities of test stars that pass within a distance $r_p\lap a$.
However there is a complicated dependence of $D_1$ on $p_{max}$ and $V$.
For $V\gg V_{bin}$, collisions at small impact parameter ($p<a$) result in the field star passing through the binary with almost no change in velocity on average, and hence $D_1\rightarrow 0$.
As $V$ is decreased, the velocity changes in the field star become greater on average and $D_1$ is determined by the distribution of angles over which field stars are ejected from the binary.
Figure 5 shows the distribution of field star velocity changes as a function of $p$ for $V=0.5 V_{bin}$.
At large impact parameters, $p\gg p_{crit}$, where
\beq
{p_{crit}\over a} \approx \sqrt{2G\m12 \over V^2 a},
\eeq
the velocity changes are distributed in the way expected for 
interaction with a point mass 
(Fig. 5a; eq. \ref{dvp}).
When $p\approx p_{crit}$, the distribution of $\delta\vf$'s broadens due to 
the change in velocity (both magnitude and direction) resulting
from interaction with the binary; 
the broadened distributions extend to positive $\delta\vf$'s, 
corresponding to stars which are ejected with high velocity in the same 
direction as their initial motion -- producing an ``anti-frictional'' 
force on the binary (Fig. 5c).
At still smaller impact parameters, $p<p_{crit}$, 
stars are ejected with a typical angle $\sim \pi/3$ with respect to 
their initial velocity vector,
compared to $\sim\pi$ for interactions with a point mass;
thus $|\overline{\delta{\vf}_x}|$ is reduced and with it the 
contribution to the frictional force (Fig. 5d).

The dynamical friction coefficient for the binary is (cf. equation \ref{df})
\beq
\df = \int\ff({\bf v}_f) \left({v-{\vf}_x\over V}\right)\langle\delta v_x\rangle d{\bf\vf}.
\eeq
We are primarily interested in the low-$v$ limit.
Repeating the analysis leading to equation (\ref{df0}) we find, for a Maxwellian $\ff$,
\beq
A = {1\over 3}\sqrt{2\over\pi} \sf^{-1} \int_0^{\infty} d\left({V\over\sf}\right)\left({V\over\sf}\right)^3 e^{-V^2/2\sf^2} \langle\delta v_x\rangle.
\eeq
This quantity was computed after scaling the velocity changes obtained from the scattering experiments, given assumed values for the two dimensionless parameters
\beq
R\equiv {p_{max}\over p_f},\ \ \ \ S\equiv {V_{bin}\over\sf},
\eeq
with $\pf=G\m12/\sf^2$; the parameter $S$ measures the hardness of the binary.
Figure 6 shows $R_1\equiv A(R,S)/\ac$, the dynamical friction coefficient for the binary normalized by the point-mass coefficient (equation \ref{df0}).
Values of $p_{max}\lap a$ are not physically interesting, hence $R$ and $S$ are restricted to $R>1/S^2$.
The greatest reduction in the frictional force occurs at small $R$ and $S$.
Small $R$ corresponds to low $p_{max}$, hence to a small contribution of distant encounters to $\df$.
Small $S$ implies a wide binary (in comparison with $p_{max}$, say) and hence a larger fraction of stars that interact strongly with the binary.
In the case $R\approx 1$, i.e. $p_{max}\approx p_f$, 
we see that the reduction in the frictional force is $\sim 50\%$ for the widest binaries, $V_{bin}\approx \sf$, falling to $\sim 20\%$ for $V_{bin}=2\sf$ and $\sim 5\%$ for $V_{bin}=5\sf$.

Results presented so far were averages over all orientations of the binary's orbital plane with respect to the initial velocity of the field star, and the direction of the binary's center-of-mass motion.
In reality, the binary will have a particular orientation with respect to 
its center-of-mass motion, 
at least on time scales short compared to the time required for the orientation of the binary or its direction of motion to change.
The typical velocity change of an interacting field star would be expected to depend systematically on the angle $\Psi$ between the normal to the binary orbital plane and ${\bf V}$.
Figure 7 illustrates the dependence $D_1(\Psi,p_{max})$ for the case 
$V=1.0\times \sqrt{G\m12/a}$.
The dynamical friction force is greatest when the binary orbital plane is perpendicular to the relative velocity vector, although the dependence on $\Psi$ is mild.
In the case of low impact parameters, $p_{max}\lap a$,  and
$\Psi\approx 0$,
the frictional force is close to zero since the field stars pass through the middle of the binary at high enough velocity that almost no deflection takes place.

\subsection{Diffusion}

We may similarly define dimensionless ratios $D_2$ and $D_3$ between the diffusion coefficients $\langle\delta v_x^2\rangle$ and $\langle\delta v_y^2+\delta v_z^2\rangle$
computed from the binary scattering experiments, 
and their equivalents for a point-mass scatterer, equations 
(\ref{scat2}) and (\ref{scat3}):
\begin{mathletters}
\begin{eqnarray}
D_2(V,p_{max}) & = & {1+p^2_{max}/p_0^2\over 4 p^2_{max}/p_0^2} \left({V\over\vb}\right)^2 \int_0^{p_{max}/a} d\left({p\over a}\right)^2 \left({{\overline{\delta v_x^2}}\over \vb}\right),  \label{d2} \\
D_3(V,p_{max}) & = & {1+p^2_{max}/p_0^2\over 4
\left[\left(1+p^2_{max}/p_0^2\right)\ln\left(1+p^2_{max}/p_0^2\right) - p^2_{max}/p_0^2\right]} \left({V\over\vb}\right)^2 \times \nonumber \\
& & \ \ \ \ \ \ \ \ 
\int_0^{p_{max}/a} d\left({p\over a}\right)^2 \left({{\overline{\delta v_y^2} + \overline{\delta v_z^2}}\over \vb}\right). \label{d3}
\end{eqnarray}
\end{mathletters}
Figures 8 and 9 plot $D_2$ and $D_3$ as functions of $p_{max}$ and $V$.
The coefficient $D_2$ describes mean square changes in the binary's velocity parallel to ${\bf V}$; these are non-dominant, i.e. bounded in their integration over impact parameter at large $p_{max}$.
Hence $D_2$ does not tend toward unity at large $p_{max}$.
At small $p_{max}$, the expected change in ${\vf}_x$ for scattering 
off a point mass is $\sim -2V$ (equation \ref{scat1}), 
whereas a typical change after interaction with the binary is $\sim -0.5\sqrt{G\m12/a}$ (cf. Fig. 5).
These are comparable when $V\approx 0.25 V_{bin}$; 
thus for $V\lap 0.25 V_{bin}$ one expects $D_2\gap 1$ for all $p_{max}$,
while for $V\gap 0.25 V_{bin}$ one expects $D_2\lap 1$ for all $p_{max}$,
in good agreement with Figure 8.
For sufficiently large $V$, $D_2$ drops to zero at small $p_{max}$ since most field stars pass through the binary without significant change in velocity.

$D_3$ describes the mean square change in the binary's velocity perpendicular to ${\bf V}$, which diverges as $\ln p_{max}$ in the case of a point-mass perturber (equation \ref{scat3}); thus $D_3\rightarrow 1$ in the limit of large $p_{max}$.
For small $p$, the change in ${\vf}_y$ for a point-mass scatterer goes to zero (cf. equation \ref{scat1}), 
while $\delta{\vf}_y^2$ remains finite in the case of interactions with the binary; hence $D_3$ diverges at small $p_{max}$.

The diffusion coefficients for the binary are
\begin{mathletters}
\begin{eqnarray}
C & = & \sqrt{2\over\pi} \int_0^{\infty} d\left({V\over\sf}\right)\left({V\over\sf}\right)^2 e^{-V^2/2\sf^2} \langle\delta v_x^2\rangle, \\
2E & = & \sqrt{2\over\pi} \int_0^{\infty} d\left({V\over\sf}\right)\left({V\over\sf}\right)^2 e^{-V^2/2\sf^2} \langle\delta v_y^2 + \delta v_z^2\rangle.
\end{eqnarray}
\end{mathletters}
Figure 10 plots the ratio $R_2=(C+2E)/(\cc+2\ec)$ between $\langle\Delta v^2\rangle$ for the binary and for a point-mass perturber in the low-$v$ limit (cf. $\S 2$).
The dashed line in that figure is
\beq
1 + {H\over 32\sqrt{2\pi}\ln\sqrt{1+2R^2}},
\label{approx2}
\eeq
the expected ratio in the limit of an infinitely hard binary
according to the approximate treatment of \S1, equation (\ref{ratio});
$\ln\Lambda$ in that expression was replaced by 
$\ln\sqrt{1+2R^2}$ (eq. \ref{approx}) and the hardening rate $H$ was set to 
$15$, appropriate for a hard binary (\cite{hil83}; \cite{qui96}).
The computed curves may be seen to tend toward the analytic expression as 
$S=V_{bin}/\sf$ increases, i.e. as the binary becomes harder.

Finally, Figure 11 gives the ratio $\sigma/\sigma_{\cal C}=(R_2/R_1)^{1/2}$, 
the factor by which the equilibrium velocity dispersion of the binary is increased relative to that of a point mass with $M=\m12$.
Based on equation (\ref{approx2}), 
this ratio is expected to be
\beq
{\sigma\over\sigma_{\cal C}} \approx\sqrt{1 + {0.18\over \ln\sqrt{1+2R^2}}},
\label{approx3}
\eeq
in the limit of a hard binary; this estimate ignores the additional
contribution to the random motion from the reduction in dynamical
friction discussed above.
Equation (\ref{approx3}) is plotted in Figure 11.
At fixed $p_{max}$, Figure 11 implies that $\sigma/\sigma_{\cal C}$ 
will increase with increasing hardness of the binary, 
until reaching a limiting value similar to equation (\ref{approx3}).
Thus the wandering of the binary should increase slightly in amplitude as 
its semi-major axis shrinks.

The results of this section may be summarized as follows.
The velocity-dependent dynamical friction coefficient $D_1(V)$
for a binary (Figure 4), with $V$ the velocity of the field star at
infinity, is typically less than that for a point particle due  
to randomization of the direction in which field stars are ejected 
during close encounters (Figure 5).
The velocity-dependent diffusion coefficient $D_3(V)$ (Figure 9) 
is typically greater than that for a point particle 
due to the gain of energy which
a field star experiences during a close encounter.
After averaging over field-star velocities,
the dynamical friction and diffusion coefficients for the binary,
$\df$ and $\langle\Delta v^2\rangle$, 
are respectively less (in absolute value) and greater than their 
values for a point particle (Figure 6 and 10).
Both effects act to increase the random velocity of a binary in 
equilibrium with field stars compared to the equilibrium random velocity of
a point particle of the same mass (Figure 11).
However the increase is modest unless $R\equiv p_{max}/\pf=p_{max}\sf^2/G\m12$ is less than about one, in other words, unless the maximum effective
impact parameter for gravitational encounters $p_{max}$ in
Chandrasekhar's theory is of order the radius of gravitational influence
of the binary $G\m12/\sf^2$.

\section{Discussion}

\subsection{Estimating $p_{max}/\pf$}

The results presented above, 
in particular equation (\ref{approx3}) and Figure 11, 
demonstrate that the random motion of a massive binary will be dominated 
by distant (elastic) encounters and therefore essentially the same as the 
Brownian motion of a point particle, unless the ratio
\beq
R\equiv {p_{max}\over\pf} = {p_{max}\sf^2\over G\m12}
\eeq
is smaller than about one.
The motivation for this result was presented already in the Introduction:
the net effect of distant (elastic) encounters is proportional to
$\ln\Lambda \approx \ln\sqrt{1+2R^2}$ (cf. equation \ref{gofr2}),
hence $R$ must be small for close (inelastic) encounters to dominate.

If the massive binary is located in a constant density region
of radius $\sim r_c$ at the center of a stellar system with a 
steeply falling density profile, 
then $p_{max}\approx r_c$ (Appendix A).

To estimate $p_f$, we note that $\pf\approx r_G$, 
the radius of gravitational influence of a body of mass $\m12$ embedded
in a stellar system with velocity dispersion $\sf$.
In most applications of Chandrasekhar's theory, 
$\pf$ is smaller than $p_{max}$ by orders of magnitude 
and $R\gg 1$.
For instance, scaling to parameters appropriate to stars in
a globular cluster,
\beq
\pf \equiv {GM\over\sf^2}\approx {G\mf\over\sf^2} \approx 
4.3\times 10^{-5} {\rm pc} 
\left({\sf\over 10\ {\rm km\ s}^{-1}}\right)^{-2} \left({\mf\over \msun}\right),
\eeq
much smaller than a core radius of $\sim 1$ pc.
However in the situations of interest here, $\pf$ is larger, 
and $R$ smaller, by roughly a factor $\m12/\mf\gg 1$ 
(ignoring changes in $\sf$).
In the case of a binary consisting of two $10\msun$ black holes 
in a globular cluster, 
we have
\beq
p_f \approx 10^{-3} {\rm pc} 
\left({\sf\over 10\ {\rm km\ s}^{-1}}\right)^{-2} 
\left({\m12\over 20 \msun}\right) .
\eeq
In the case of a supermassive black hole binary at the center of a galaxy,
we can use the $M_{\bullet}-\sigma$ relation
\beq
\m12 =M_{\bullet} \approx 1.30\times 10^8\msun \left({\sf\over 200\ {\rm km\ s}^{-1}}\right)^{4.7}
\label{msig}
\eeq
(\cite{mef01})
to write
\beq
p_f \approx 10\: {\rm pc} 
\left({\sf\over 200\ {\rm km\ s}^{-1}}\right)^{2.7}.
\eeq

Consider first a binary black hole at the center of a globular cluster.
A core radius of $r_c\approx 1$ pc is still much
greater than $p_f\approx 10^{-3}$ pc, 
implying $R\gg 1$ and no appreciable enhancement of the Brownian motion.
However in the ``post-core-collapse'' globular clusters
(\cite{djk86}) the stellar density follows a steep power law all the way 
into the observable center and $r_c$ is much smaller than 1 pc.
While $p_{max}$ is not well defined in this case,
arguments like those in the Appendix suggest that 
the effective $p_{max}$ would be small, 
hence $R\lap 1$ and a substantial enhancement in the Brownian 
motion is predicted (Figure 11).

Next consider the case of a $\sim 10^8\msun$ black-hole binary in a
galactic nucleus.
Like globular clusters, some galaxy nuclei have ``cores''
(\cite{lau95}), 
actually weak power laws in the space density (\cite{mef96}).
(The cores of globular clusters could also be weak power laws.)
The ``break radius'' $r_b$ in these galaxies (\cite{fab97}) 
plays roughly the role of a core radius;
Poon \& Merritt (2000) note that $r_b\approx \mathrm{a\ few} \times\: r_G$ 
in two galaxies where both $r_b$ and $M_{\bullet}$ can be accurately 
measured, hence $p_{max}\gap p_f$ and $R\gap 1$.
In the ``power-law'' galaxies (\cite{fer94}; \cite{lau95}), 
which have no detectable core, $R$ would be smaller, $R\lap 1$.
The enhancement of the Brownian motion could be substantial 
for a massive binary at the center of a power-law galaxies,
perhaps as much as a factor of two in $\sigma$ (Figure 11).

The concept of a maximum effective impact parameter
$p_{max}$ on which these estimates depend is a loose one, 
both as used here and in Chandrasekhar's original theory.
Refining these estimates of $p_{max}$ will require $N$-body simulations.

\subsection{Equilibration time scales}

Here we test the assumption made above (\S 2) that
the center-of-mass motion of the binary reaches a statistical 
steady state in a time short compared to the time over which the
orbital elements of the binary change.
In fact this is not always the case.
The time scale for change of the binary's semi-major axis 
(cf. equation \ref{def_h}) is
\beq
t_{\mathrm harden} \equiv \left[ a{d\over dt}\left({1\over a}\right)\right]^{-1} = {\sf\over G\rho_f a H}.
\eeq
Approach to statistical equilibrium of the Brownian motion occurs in a time
(cf. equation \ref{123})
\beq
t_{\mathrm harden} \equiv \left|{v\over\df}\right| \approx
\left|{v^2\over\langle\Delta v^2\rangle}\right| \approx {1\over A} \approx {1\over A_{\cal C}} =
{3\over 4\sqrt{2\pi}}{\sf^3\over G^2\m12\rho_f\ln\Lambda}.
\eeq
The ratio is
\beq
{t_{\mathrm harden}\over t_{\mathrm relax}} = 
{4\sqrt{2\pi}\over 3H}{G\m12\over a\sf^2}\ln\Lambda = 
{4\sqrt{2\pi}S^2\ln\Lambda\over 3H} \approx
{3.3S^2\ln\sqrt{1+R^2}\over H}
\eeq
with $S^2\equiv V_{\mathrm bin}^2/\sf^2=G\m12/\sf^2a$.
This ratio must be large if the binary's semi-major axis is to remain
nearly fixed during the time required for the Brownian motion to be
established.
For large enough $R$, i.e. large enough $p_{max}$, this is guaranteed.
However the case of most interest here is $R\approx 1$ (\S 5.1), for which
\beq
{t_{\mathrm harden}\over t_{\mathrm relax}} \approx {S^2\over H}.
\eeq
For a very hard binary, $H\approx 15$ (\cite{qui96}) and
$t_{\mathrm decay}/t_{\mathrm relax}\approx 0.07 V^2_{\rm bin}/\sf^2$,
which exceeds unity for $S=V_{\mathrm bin}/\sf\gap 4$.
Thus the condition is satisfied for sufficiently hard binaries.
For a soft binary, $S\lap 1$, we can use the 
expressions in Quinlan (1966, eqs. 16-18),
to write $H\approx 4.4 S^2$, giving 
$t_{\mathrm decay}/t_{\mathrm relax}\approx 0.23$.
Thus in the case of a soft binary,
the binary separation {\it will} change appreciably
during the time that the Brownian motion is being established.
However in the soft regime, the Brownian velocity dispersion
depends only very weakly on $S$ (cf. Figure 11), so
the conclusions presented above are not significantly affected.

\subsection{Consequences of Brownian motion}

The Brownian velocity dispersion of the binary is predicted to be 
(equation \ref{eq_equi})
\beq
\langle v^2\rangle = 3\sigma^2 = 3{R_2\over R_1}{\mf\over\m12}\sf^2.
\label{eq_equi2}
\eeq
We can convert this into a mean square displacement by assuming that 
the binary moves in a harmonic potential well produced by a uniform density 
core of field stars.
The virial theorem
\beq
\langle v^2\rangle = -\langle {\bf F} \cdot {\bf r}\rangle
\eeq
as applied to the binary's center-of-mass motion relates the
mean square velocity of the binary to an average
of the force acting on it.
In a spherical constant-density core, the force (neglecting encounters) is
$-(4\pi G\rf/3){\bf r}$
and
\beq
\langle v^2\rangle = {4\over 3} \pi G\rf \langle r^2\rangle.
\eeq
Writing $r_c^2=9\sf^2/4\pi G\rf$ (\cite{kin66}) yields
\beq
{\langle r^2\rangle\over r_c^2} = {R_2\over R_1}{\mf\over\m12}.
\label{wander1}
\eeq
This treatment (cf. \cite{baw76}) is approximate in that it 
ignores the effect of the motion of the binary on the background stars
and assumes that gravitational encounters are uncoupled from the
quasi-periodic motion in the core.

For a binary black hole at the center of a globular cluster,
the rms Brownian velocity is
\beq
v_{rms} \approx 3.87\: {\rm km\ } {\rm s}^{-1}\left({R_2\over R_1}\right)^{1/2} 
\left({20\mf\over\m12}\right)^{1/2} \left({\sf\over 10\: {\rm km\ s^{-1}}}\right) 
\eeq
and the wandering radius is
\beq
r_{rms} \approx 0.22\: {\rm pc} \left({R_2\over R_1}\right)^{1/2}
\left({20\mf\over\m12}\right)^{1/2}
\left({r_c\over 1\ {\rm pc}}\right).
\label{wander2}
\eeq

In the case of a binary supermassive black hole in a galactic nucleus,
the Brownian velocity is only
\beq
v_{rms} \approx 0.030\: {\rm km\ } {\rm s}^{-1} \left({R_2\over R_1}\right)^{1/2}\left({\mf\over\msun}\right)^{1/2} \left({\sf\over 200\: {\rm km\ s^{-1}}}\right)^{-1.35} 
\eeq
and the wandering radius is
\beq
r_{rms} \approx 0.0087 \: {\rm pc} \left({R_2\over R_1}\right)^{1/2}\left({\sf\over 200\: {\rm km\ s}^{-1}}\right)^{-2.35}\left({r_c\over 100\ {\rm pc}}\right);
\label{wander}
\eeq
the $M_{\bullet}-\sigma$ relation (\ref{msig}) 
has again been used to relate $\m12$ to $\sf$.
The Brownian motion of a supermassive black hole is of course
very small, even with the enhancement due to inelastic scattering
derived here.
The predicted wandering radius is smaller than the likely
separation of the black holes in a binary system; in fact it is of 
order the separation at which gravitational radiation would lead to a rapid 
coalescence (\cite{mer00}).

Measurement of the velocity of the Milky Way black hole
has recently become feasible (\cite{rei99}; \cite{bas99}).
The upper limit of $\sim 20\ $\kms is a factor $\sim 200$ greater
than the expected Brownian velocity given its mass of 
$\sim 3\times 10^6\msun$.

The wandering amplitude of a binary supermassive black hole
in a galactic nucleus has recently been discussed by a number of authors
(\cite{mak97}; \cite{quh97}; \cite{mer00}; \cite{gor00}).
The problem is interesting because a stationary black-hole
binary will quickly eject all of the stars with pericenters
less than $\sim 2a$ (e.g. \cite{zie00}); once this occurs,
the density of stars around the binary drops and the decay stalls.
Wandering is a possible way to increase the number of stars that a binary
can interact with, thus allowing the binary to
decay to the point that gravitational radiation coalescence
can occur (\cite{pet64}).

Quinlan \& Hernqust (1997), in a series of computer simulations
using a hybrid $N$-body code,
noticed a wandering of a massive binary with an amplitude
more than five times greater than expected on the basis of an equation like
(\ref{wander}).
The core radius that they cite, $\sim 0.04$, is comparable to $r_G$ in
their simulations, so $R\approx r_c/r_G\approx 1$.
Figure 11 suggests that the increase in the amplitude of the 
wandering would probably be much less than the claimed factor of $5-10$.

Makino (1997) carried out $N$-body merger simulations similar to those of
Quinlan \& Hernquist (1997) but using a more conservative,
direct-summation code.
Makino's Figure 7 shows a wandering amplitude that scales as $\sim N^{-1/2}$;
the rms velocity of the binary in that plot is 
comparable, or perhaps $\sim 50\%$ larger, than expected for a point mass, 
consistent with the results obtained here.

Quinlan \& Hernquist (1997) used a spectrum of masses for the field particles,
spanning a range of $\sim 10^3$ in $\mf$, and adopted the lowest mass
when computing the expected amplitude of the binary's Brownian motion.
Mass segregation might have brought some of the more massive
particles into the nucleus over the course of their simulations.
Quinlan \& Hernquist also used a basis-function-expansion code
for computing the forces between field stars.
Such codes do not conserve momentum and this too may have contributed
to the random motion of the binary.
A repeat of the Quinlan \& Hernquist simulations using a 
different $N$-body code could help to clarify the discrepancy
between their results and those of Makino.

If supermassive black hole binaries wander as little as implied by
equations (\ref{approx3}) or (\ref{wander}),
it is difficult to see how they could interact with enough stars to 
allow them to achieve gravitational-radiation coalescence in a reasonable 
time (\cite{mer00}).
Unless some other mechanism is effective at removing energy from these
binaries, one might expect them to persist for the lifetime of a galaxy
at separations $a \lap 0.1$ pc,
or at least until a subsequent merger event brings another supermassive
black hole into the nucleus (\cite{val96}).

\acknowledgments
I thank Milo\v s Milosavljevi\' c for advice about evaluating 
some of the integrals and for numerous discussions about the 
ideas presented here.
A conversation with Douglas Heggie helped to motivate this paper.
This work was supported by NSF grants AST 96-17088 and 00-71099 and
by NASA grants NAG5-6037 and NAG5-9046.

\appendix
\section{}

Maoz (1993, eq. 4.4) gives an implicit expresson for $p_{max}$ 
appropriate for an object at the center of a spherically-symmetric 
matter distribution $\rho(r)$.
Maoz's formula,
\beq
{1\over\rho(0)}\int_{p_{min}}^{\infty} {\rho(r)\over r} dr,
\label{maoz1}
\eeq
replaces the Coulomb logarithm in expressions like equation (\ref{ac}).
If we set
\beq
\rho(r) = \rho(0) \left(1+{r^2\over r_c^2}\right)^{-\gamma/2},
\eeq
with $r_c$ the core radius of the field stars,
Maoz's formula gives
\beq
\ln\Lambda \approx 	\left\{ \begin{array}{ll}
		\sinh^{-1}\left({r_c\over p_{min}}\right) & \mbox{if $\gamma=1$,} \\
		{1\over 2}\ln\left({p^2_{min}+r_c^2\over p^2_{min}}\right) & 
		\mbox{if $\gamma=2$.}
		\end{array}
		\right.
		\label{cases2}
\eeq
Setting $p_{min}\approx \pf \approx G\m12/\sf^2$ (eq. \ref{pmin}) and equating $\ln\Lambda$ with $\ln\sqrt{1+2R^2}$ (eq. \ref{approx}), we find
\beq
R \approx {r_c\over\pf} \approx {r_c\over r_G}
\eeq
for both values of $\gamma$.

\clearpage

\begin{figure}
\plotone{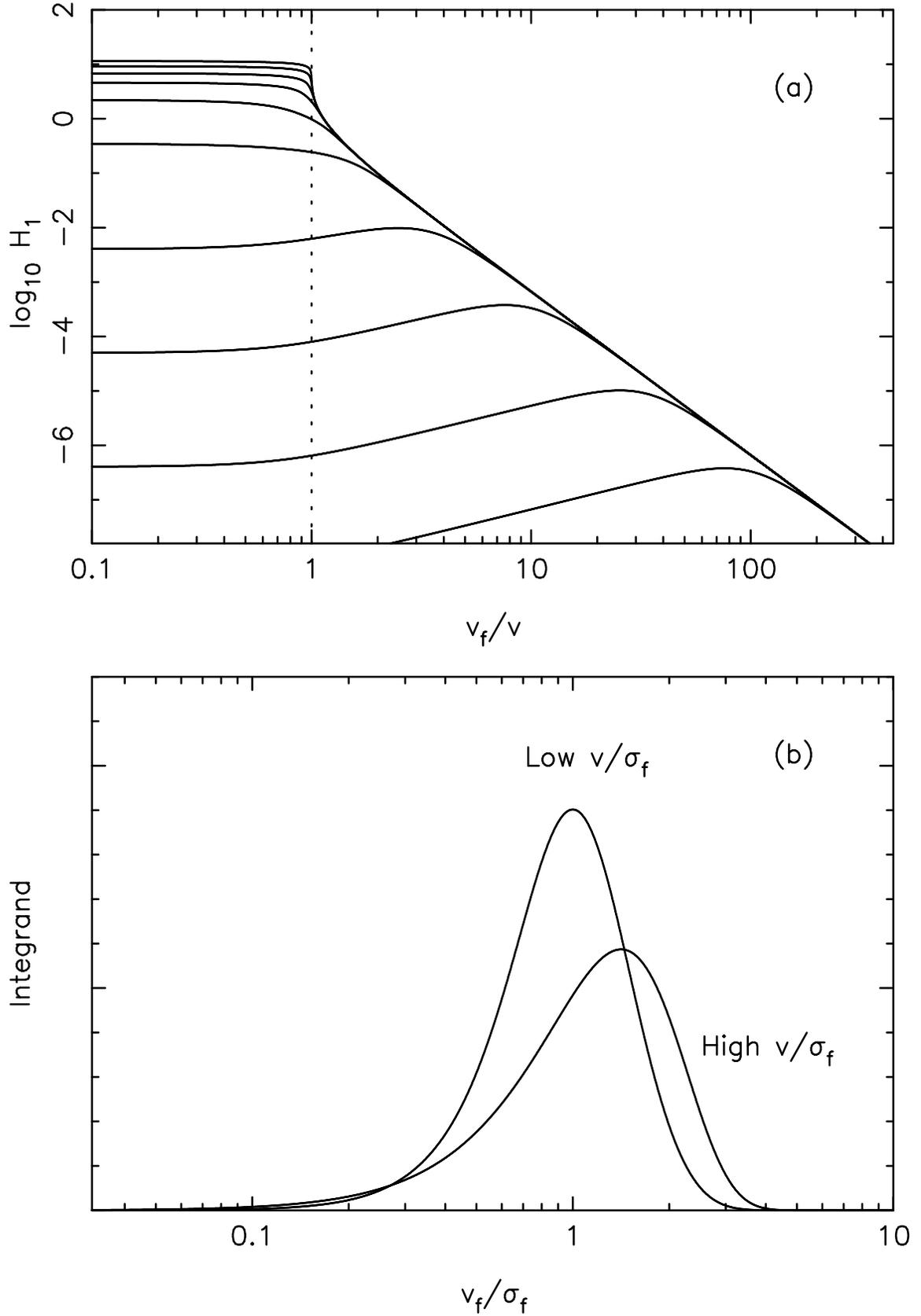} 
\caption{
(a) $H_1(v,v_f,p_{max})$, equation (\ref{hofv}), 
for $p_{max}/p_f=1$ and $v/\sigma_f=\{0.01,0.03,0.1,0.3,1,3,10,30,$ $100,300\}$; $v/\sf$ increases upwards.
(b) The relative contribution of different field-star velocities to the dynamical friction integral (\ref{dfh1}), in the limit of low and high test-star velocities.
}

\end{figure}

\begin{figure}
\plotone{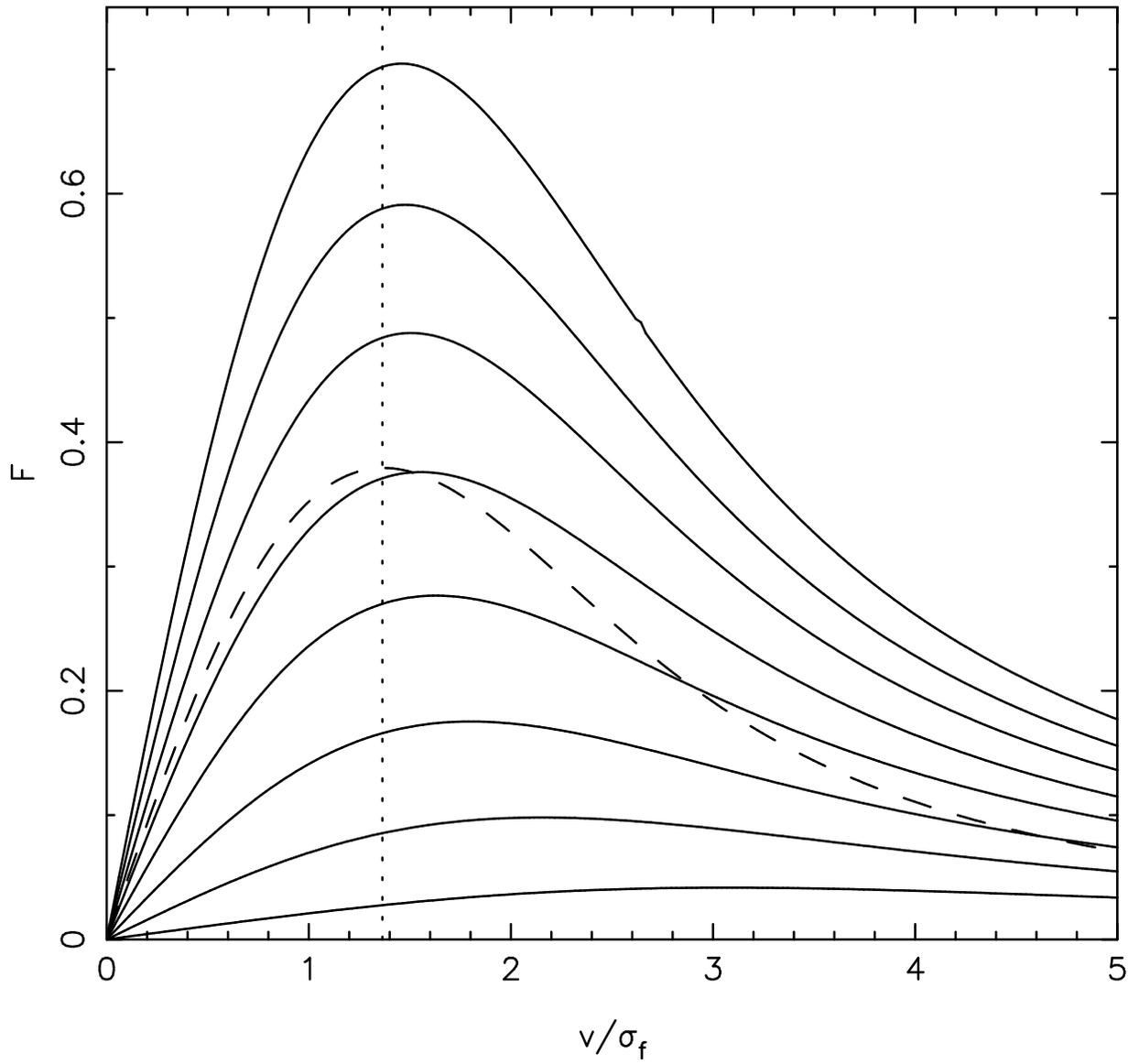} 
\caption{
The dimensionless dynamical friction coefficient $F(v,p_{max})$ (equation \ref{coef2}) for $p_{max}/p_f = \{0.3,1,3,10,30,100,300,1000\}$.
Dashed curve is the classical expression with $\ln\Lambda=4$;
vertical dotted line indicates the position of the maximum of this function.
}
\end{figure}

\begin{figure}
\plotone{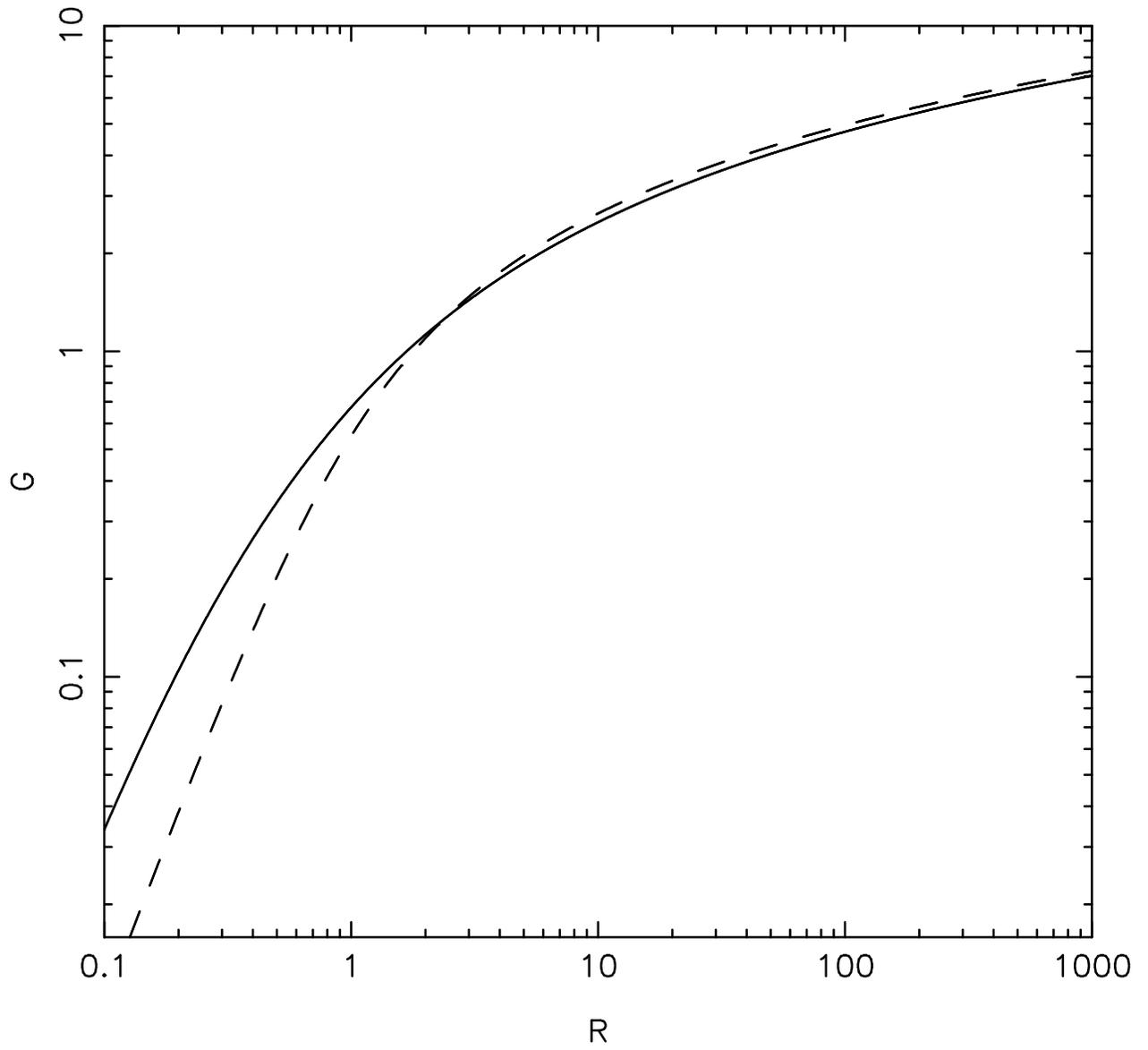} 
\caption{
The function $G(R)$ (equation \ref{gofr}).
The dashed line is the approximation of equation (\ref{gofr2}),
$G(R)\approx\ln\sqrt{1+2R^2}$.
}
\end{figure}

\begin{figure}
\plotone{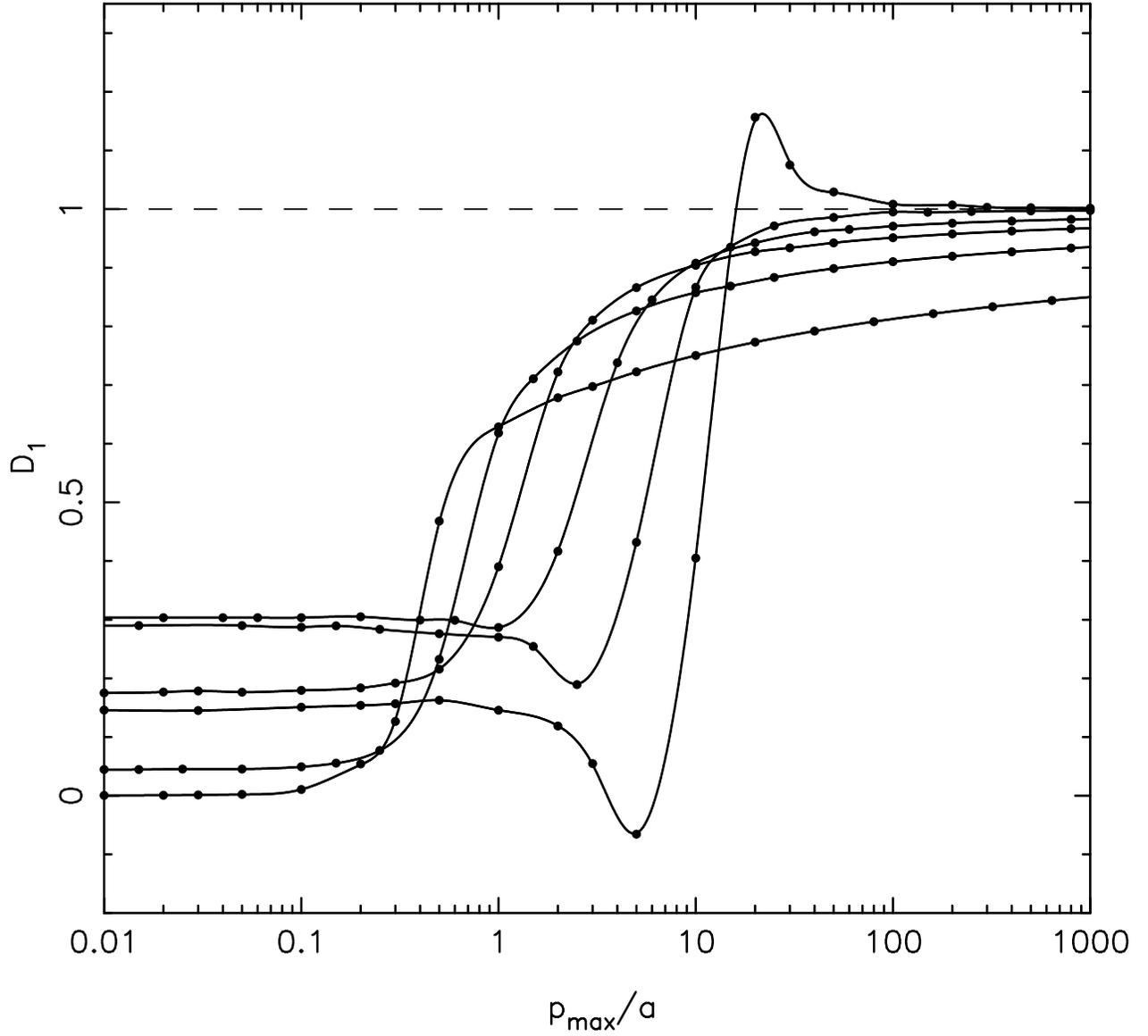} 
\caption{
Dynamical friction reduction factor $D_1(V,p_{max})$ (eq. \ref{d1}).
$D_1$ is defined as the mean change in field-star velocities after 
interaction with the binary,
expressed as a fraction of the value expected for interactions with a point-mass scatterer of the same total mass as the binary;
 $p_{max}$ is the maximum impact parameter and $a$ is the binary semi-major axis.
Different curves correspond to relative velocities at infinity of
$V/V_{bin} = 0.1,0.2,0.5,1,2,10$; 
the lowest (highest) $V$ produces the highest (lowest) $D_1$ at large $p_{max}$.
Points are averages computed from the numerical integrations; curves are spline fits.
Some of the curves have been extended to large $p_{max}/a$ using the analytic expression for a point-mass scatterer.
}
\end{figure}

\begin{figure}
\plotone{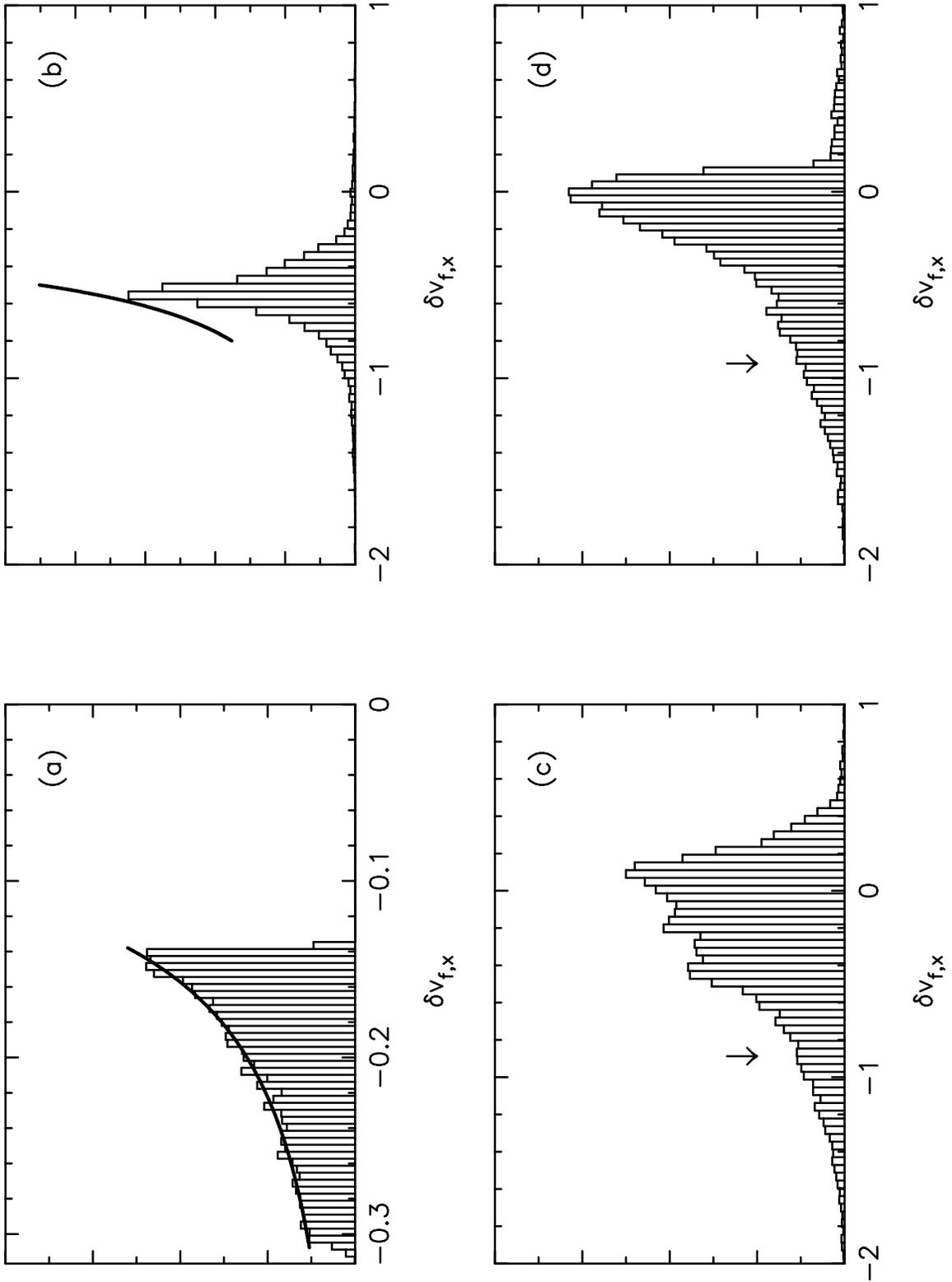} 
\caption{Distribution of field star velocity changes for scattering experiments with $V/V_{bin}=0.5$.
Each plot corresponds to $5\times 10^4$ scattering experiments within some range of impact parameters $[p_1,p_2]$ in units of $a$. 
(a) $[6,10]$
(b) $[2,4]$
(c) $[0.6,1]$
(d) $[0.4,0.6]$.
Solid lines in (a) and (b) are the distributions corresponding to 
scattering off a point-mass perturber.
In (c) and (d), the mean of this distribution (which is very narrow) 
is indicated by the arrows.
The magnitude of the frictional force is proportional to the mean
value of $-\delta{\vf}_x$; 
at low impact parameters, the distribution of $\delta{\vf}_x$'s
is shifted toward zero and the frictional force is reduced.}
\end{figure}

\begin{figure}
\plotone{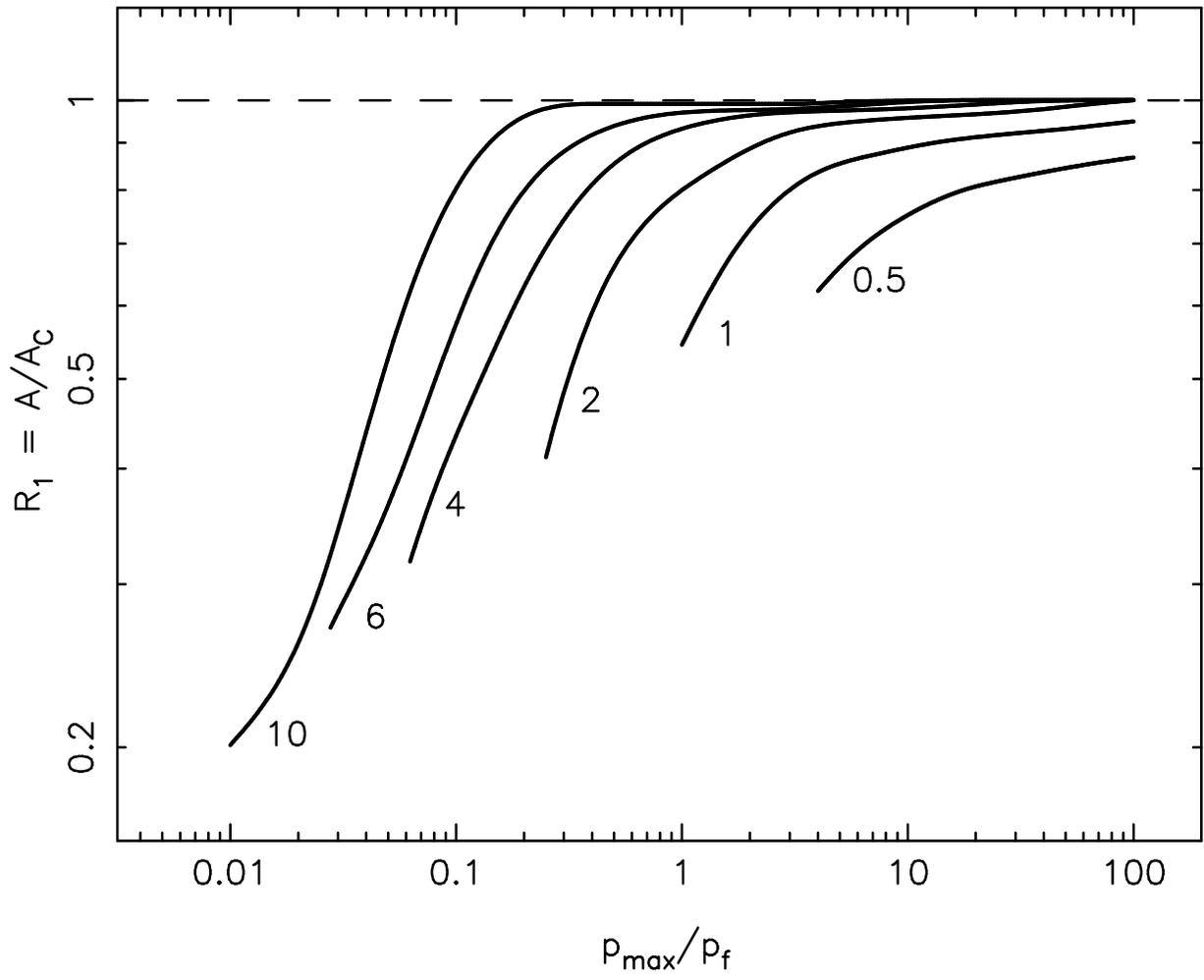} 
\caption{Reduction in the dynamical friction force for a massive binary moving at low velocity, $v\ll\sigma_f$.
Different curves correspond to $V_{bin}/\sigma_f= 0.5,1,2,5,10$;
the lowest (highest) $V_{bin}$ produces the the highest (lowest) reduction factor at large $p_{max}/p_f$.
}
\end{figure}

\begin{figure}
\plotone{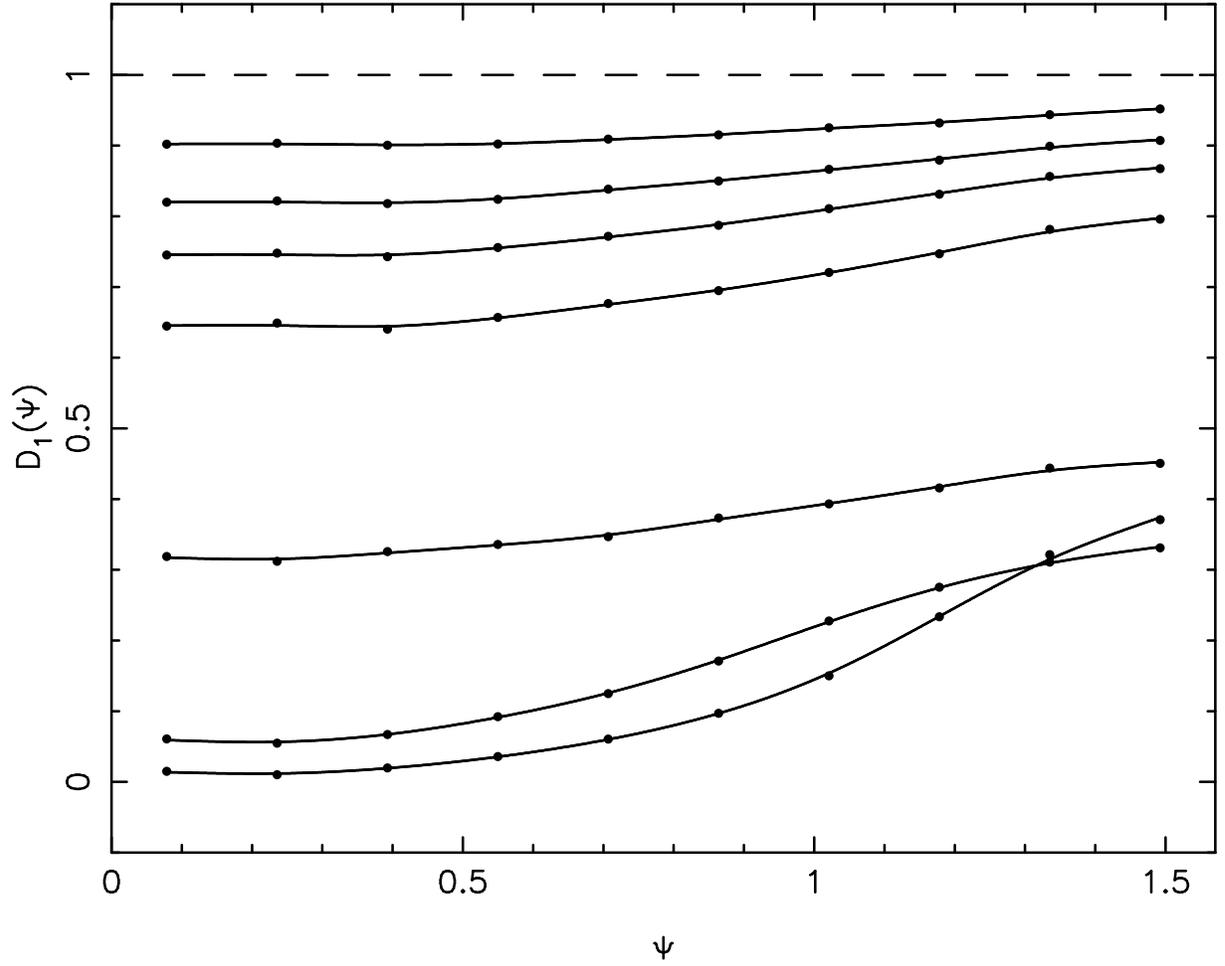} 
\caption{
Dynamical friction reduction factor as a function of angle $\Psi$ between the normal to the binary's orbital plane and the relative velocity vector of the field stars, for $V/V_{bin} = 1$.
Different curves correspond to different values of the maximum impact parameter; proceeding upward, $p_{max}/a = (0.2,0.5,1,2,3,5,20)$. 
Points are averages computed from the numerical integrations; curves are spline fits.
}
\end{figure}

\begin{figure}
\plotone{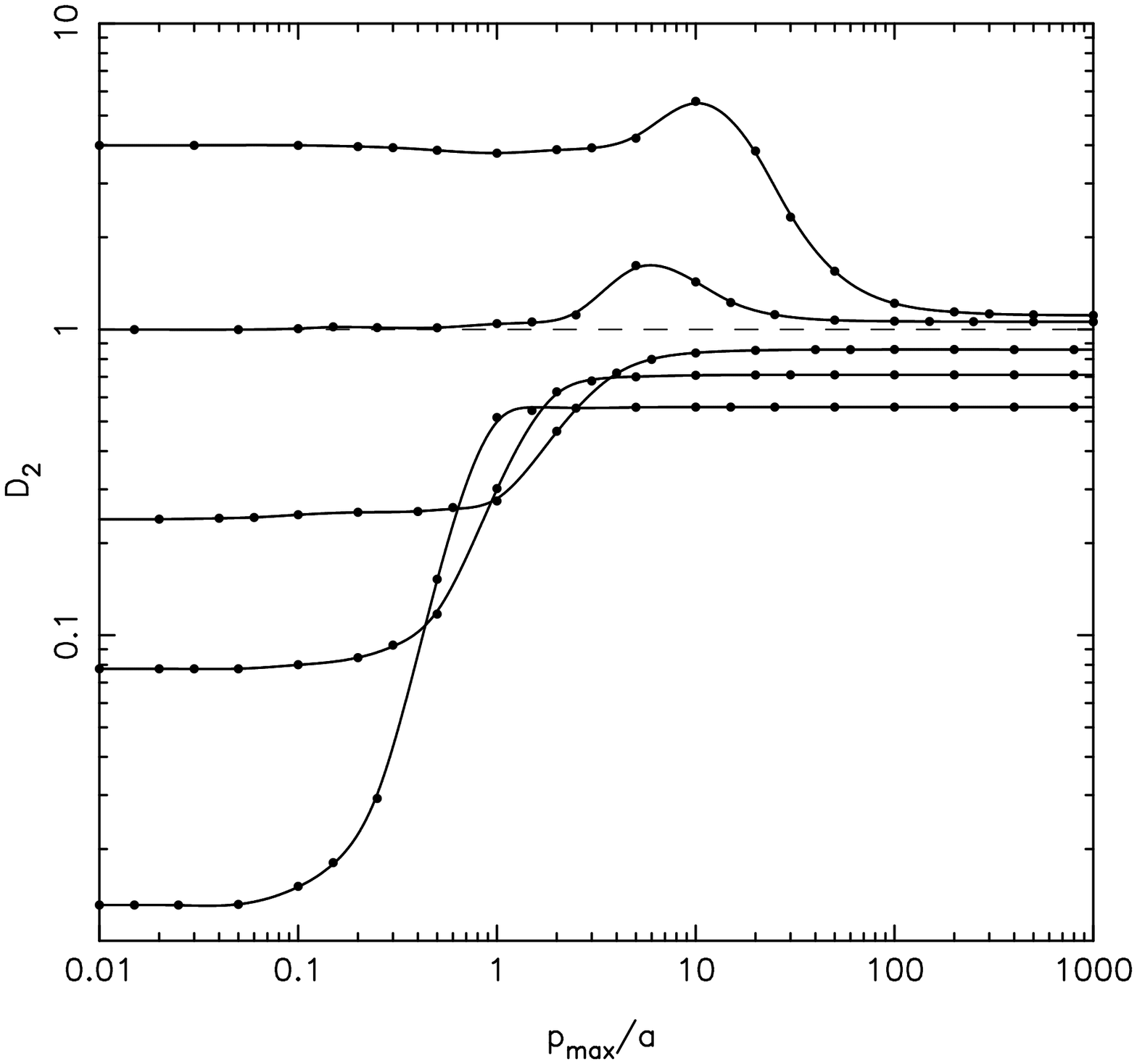} 
\caption{
Dimensionless ratio $D_2(V,p_{max})$ (eq. \ref{d2}) 
describing changes in the mean square field-star velocity in a direction parallel to the initial velocity, normalized to the value for a point-mass scatterer.
Different curves correspond to $V/V_{bin} = (0.1,0.2,0.5,1,2)$; 
the lowest (highest) $V$ produces the highest (lowest) $D_2$ at small $p_{max}$.
}
\end{figure}

\begin{figure}
\plotone{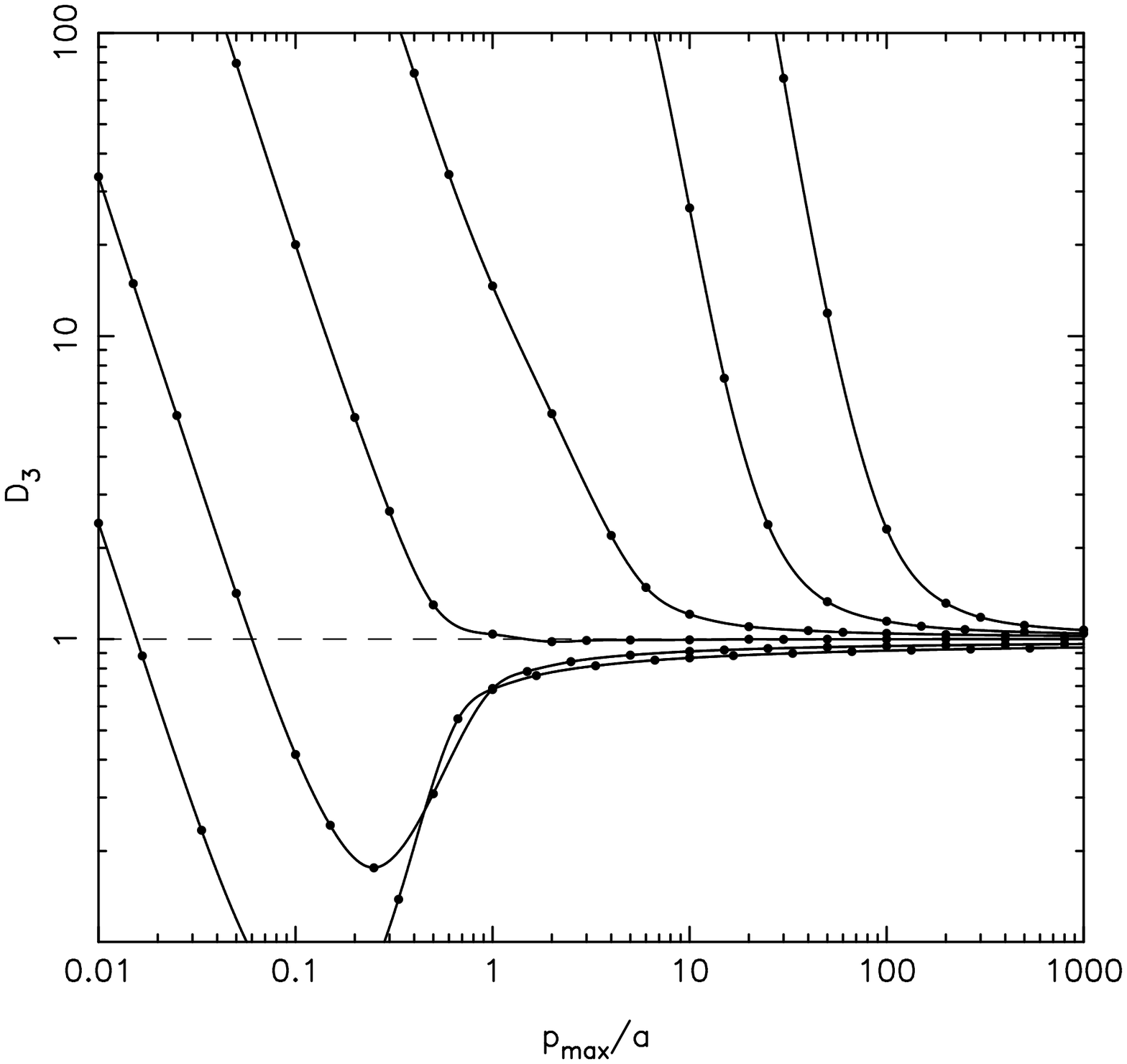}
\caption{
Dimensionless ratio $D_3(V,p_{max})$ (eq. \ref{d3}) 
describing changes in the mean square field-star velocity in a direction perpendicular to the initial velocity, normalized to the value for a point-mass scatterer.
Different curves correspond to $V/V_{bin} = 0.1,0.2,0.5,1,2,3$; 
the lowest (highest) $V$ produces the highest (lowest) $D_3$ at small $p_{max}$.
}
\end{figure}

\begin{figure}
\plotone{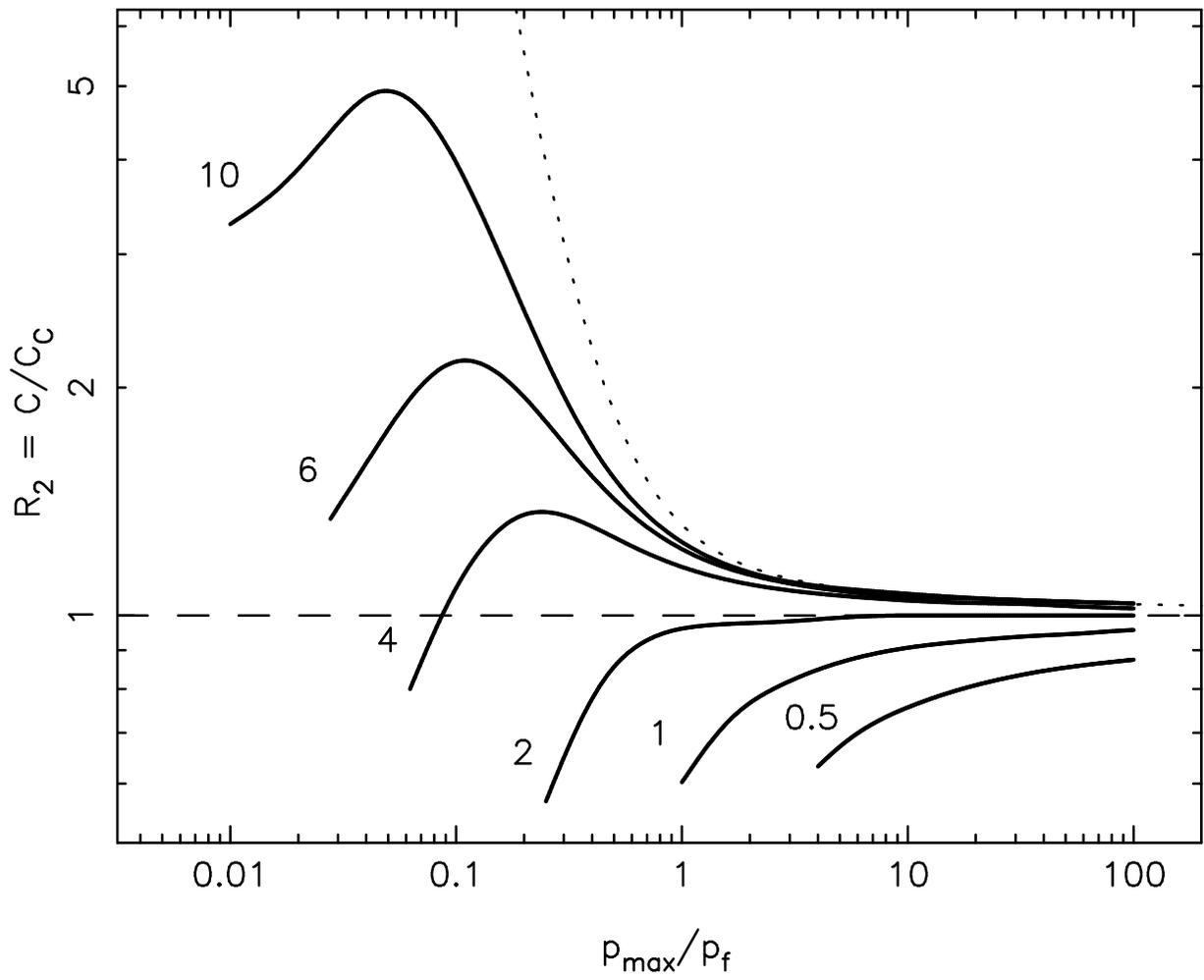} 
\caption{Increase in the diffusion coefficient $\langle\Delta v^2\rangle$ 
for a massive binary moving at low velocity, 
$v\ll\sf$, compared to its value for a point mass.
Curves are labelled by the value of $V_{bin}/\sigma_f$.
Dashed line is the approximation of equation (\ref{approx2}).
}
\end{figure}

\begin{figure}
\plotone{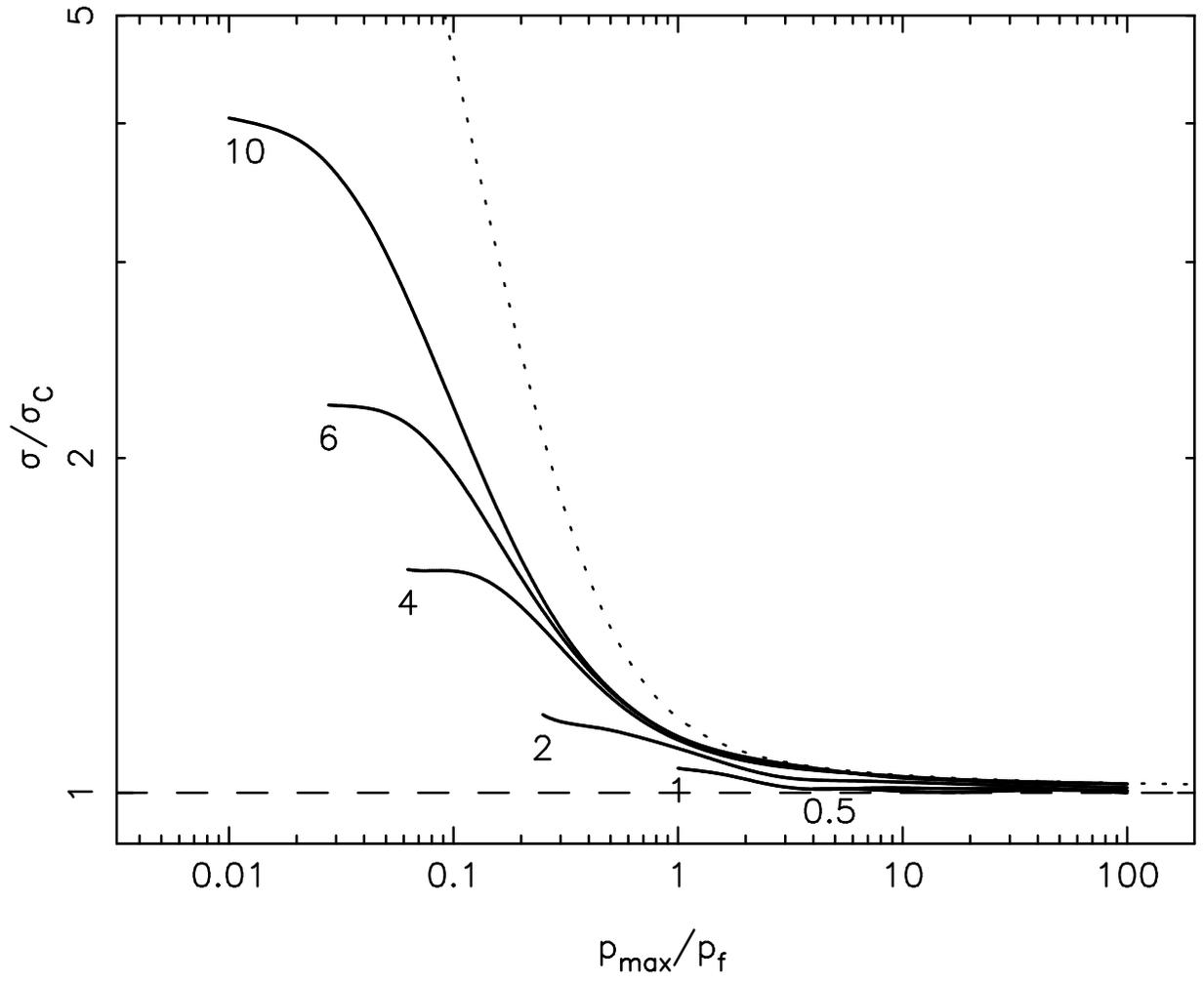} 
\caption{Increase in the equilibrium velocity dispersion of a massive binary compared to its value for a point mass.
Curves are labelled by the value of $V_{bin}/\sf$.
Dashed line is the approximation of equation (\ref{approx3}).
}
\end{figure}
\end{document}